\documentclass[12pt,preprint]{aastex}
\newcommand  \kms      {\ifmmode {\rm km\,s}^{-1} \else km\,s$^{-1}$\fi}

\newcommand  \ergs     {\ifmmode {\rm ergs\,s}^{-1} \else ergs s$^{-1}$\fi}
\newcommand  \ergcms   {\ifmmode {\rm ergs\,cm}^{-2}\,{\rm s}^{-1}
                        \else ergs\,cm$^{-2}$\,s$^{-1}$\fi}
\newcommand  \ergcmsA {\ifmmode{\rm ergs\,cm}^{-2}\,{\rm s}^{-1}\,{\rm\AA}^{-1}
                        \else ergs\,cm$^{-2}$\,s$^{-1}$\,\AA$^{-1}$\fi}
\newcommand \ergcmsHz {\ifmmode{\rm ergs\,cm}^{-2}\,{\rm s}^{-1}\,{\rm Hz}^{-1}
                        \else ergs\,cm$^{-2}$\,s$^{-1}$\,Hz$^{-1}$\fi}
\newcommand  \phcms    {\ifmmode {\rm ph\,cm}^{-2}\,{\rm s}^{-1}
                        \else ,ph\,cm$^{-2}$\,s$^{-1}$\fi}
\newcommand  \phcmsA   {\ifmmode {\rm ph\,cm}^{-2}\,{\rm s}^{-1}\,{\rm\AA}^{-1}
                        \else ph\,cm$^{-2}$\,s$^{-1}$\,\AA$^{-1}$\fi}
\def\micron{\ifmmode \mu{\rm m} \else $\mu$m\fi}
\def\kms{\ifmmode {\rm km\,s}^{-1} \else km\,s$^{-1}$\fi}
\def\Hubble{\ifmmode {\rm km\,s}^{-1}\,{\rm Mpc}^{-1}
        \else km\,s$^{-1}$\,Mpc$^{-1}$\fi}
\def\ergsec{\ifmmode {\rm ergs\;s}^{-1} \else ergs s$^{-1}$\fi}
\def\ergscm{\ifmmode {\rm ergs\,s}^{-1}\,{\rm cm}^{-2}
          \else ergs\,s$^{-1}$\,cm$^{-2}$\fi}
\def\ergscmA{\ifmmode {\rm ergs\,s}^{-1}\,{\rm cm}^{-2}\,{\rm \AA}^{-1}
          \else ergs\,s$^{-1}$\,cm$^{-2}$\,\AA$^{-1}$\fi}
\def\ergscmHz{\ifmmode {\rm ergs\,s}^{-1}\,{\rm cm}^{-2}\,{\rm Hz}^{-1}
q          \else ergs\,s$^{-1}$\,cm$^{-2}$\,Hz$^{-1}$\fi}
\def\Msun{\ifmmode M_{\odot} \else $M_{\odot}$\fi}
\def\Lsun{\ifmmode L_{\odot} \else $L_{\odot}$\fi}
\def\qo{\ifmmode q_{0} \else $q_{0}$\fi}
\def\Ho{\ifmmode H_{0} \else $H_{0}$\fi}
\def\ho{\ifmmode h_{0} \else $h_{0}$\fi}
\def\qo{\ifmmode q_{0} \else $q_{0}$\fi}
\def\ao{\ifmmode a_{0} \else $a_{0}$\fi}
\def\to{\ifmmode t_{0} \else $t_{0}$\fi}
\def\Halpha{\ifmmode {\rm H}\alpha \else H$\alpha$\fi}
\def\Hbeta{\ifmmode {\rm H}\beta \else H$\beta$\fi}
\def\hb{\ifmmode {\rm H}\beta \else H$\beta$\fi}
\def\Hgamma{\ifmmode {\rm H}\gamma \else H$\gamma$\fi}
\def\Hdelta{\ifmmode {\rm H}\delta \else H$\delta$\fi}
\def\Lya{\ifmmode {\rm Ly}\alpha \else Ly$\alpha$\fi}
\def\Lyb{\ifmmode {\rm Ly}\beta \else Ly$\beta$\fi}
\def\hi{\ifmmode \mbox{{\rm H}\,{\sc i}} \else H\,{\sc i}\fi}
\def\hii{H\,{\sc ii}}

\def\ciii{\ifmmode {\rm C}\,{\sc iii} \else C\,{\sc iii}\fi}

\def\oii{[O\,{\sc ii}]\,$\lambda3727$}
\def\oiii{[O\,{\sc iii}]\,$\lambda5007$}

\def\o5007{[O\,{\sc iii}]\,$\lambda5007$}
\def\oivIR  {[O\,{\sc iv}]\,$25.9 \mu$m}
\def\nevIR {[Ne\,{\sc v}]\,$14.3 \mu$m}
\def\pahIR {PAH\,$7.7 \mu$m}

\def\ne212m {[Ne\,{\sc ii}]\,$12.8 \mu$m}
\def\nulnu60 {$\nu$L$_\nu$(60$\mu$m)}

\def  \kms         {\hbox{km s$^{-1}$}}          
\def  \ergs        {\hbox{erg s$^{-1}$}}              

\def  \La          {\ifmmode {\rm Ly}\alpha \else Ly$\alpha$\fi}
\def  \Ka          {\ifmmode {\rm K}\alpha \else K$\alpha$\fi}
\def  \Lb          {\ifmmode {\rm L}\beta \else L$\beta$\fi}
\def  \Ha          {\ifmmode {\rm H}\alpha \else H$\alpha$\fi}
\def  \Hb          {\ifmmode {\rm H}\beta \else H$\beta$\fi}
\def  \Pa          {\ifmmode {\rm P}\alpha \else P$\alpha$\fi}
\def  \CIIIb       {\ifmmode {\rm C}\,{\sc iii]}\,\lambda1909
                     \else C\,{\sc iii]}\,$\lambda1909$\fi}
\def  \CIV         {\ifmmode {\rm C}\,{\sc iv}\,\lambda1549
                     \else C\,{\sc iv}\,$\lambda1549$\fi}
\def  \MgII         {\ifmmode {\rm Mg}\,{\sc ii}\,\lambda2798
                     \else Mg\,{\sc ii}\,$\lambda2798$\fi}
\def  \OVI         {\ifmmode {\rm O}\,{\sc vi}\,\lambda1035
x
                     \else O\,{\sc vi}\,$\lambda1035$\fi}

\def \spitzer      {{\it Spitzer}}
\def \KAO      {{\it KAO}}
\def \ISO      {{\it ISO}}
\def \IRAS      {{\it IRAS}}
\def \IRS      {{\it IRS}}

\shorttitle{Origin of the far-infrared continuum of QSOs}
\shortauthors{Schweitzer et al.}

\begin{document}

\title{\spitzer\ Quasar and ULIRG evolution study (QUEST):\\
 I. The origin of the far infrared continuum of QSOs }

\author{M. Schweitzer, D. Lutz, E. Sturm, A. Contursi,
L.J. Tacconi, M.D. Lehnert,  K. Dasyra, R. Genzel}
\affil{Max-Planck-Institut f\"ur extraterrestrische Physik\\
       Postfach 1312, 85741 Garching, Germany
\email{schweitzer@mpe.mpg.de, lutz@mpe.mpg.de, sturm@mpe.mpg.de,
contursi@mpe.mpg.de, linda@mpe.mpg.de, mlehnert@mpe.mpg.de, dasyra@mpe.mpg.de,
genzel@mpe.mpg.de}}

\author{S. Veilleux, D. Rupke, D.-C. Kim, A.J. Baker\altaffilmark{1}}
\affil{Department of Astronomy, University of Maryland\\ 
      College Park, MD 20742-2421, USA
\email{veilleux@astro.umd.edu, drupke@astro.umd.edu, dckim@astro.umd.edu,
ajb@astro.umd.edu}}

\author{H. Netzer, A. Sternberg}
\affil{School of Physics and Astronomy and the Wise Observatory\\
      The Raymond and Beverly Sackler Faculty of Exact Sciences\\
      Tel-Aviv University, Tel-Aviv 69978, Israel
\email{netzer@wise1.tau.ac.il, amiel@wise.tau.ac.il}}

\author{J. Mazzarella, S. Lord}
\affil{IPAC, California Institute of Technology\\
MS 100-22, Pasadena, CA 91125, USA
\email{mazz@ipac.caltech.edu, lord@ipac.caltech.edu}}

\altaffiltext{1}{Jansky Fellow, National Radio Astronomy Observatory}

\begin{abstract}
This paper addresses the origin of the far-infrared (FIR) continuum of 
QSOs, based on the Quasar and ULIRG Evolution Study (QUEST) of nearby 
QSOs and ULIRGs using observations with the {\it Spitzer Space Telescope}. 
For 27 Palomar-Green QSOs at z$\lesssim$0.3, we derive luminosities of 
diagnostic lines (\ne212m , \nevIR, \oivIR ) and 
emission features (\pahIR\ emission which is related to star formation), as 
well as 
continuum luminosities over a range of mid- to far-infrared wavelengths
between 6 and 60$\mu$m. We detect star-formation related PAH emission in 
11/26 QSOs and 
fine-structure line emission in all of them, often in multiple lines. The 
detection of PAHs in the average spectrum of sources which lack individual 
PAH detections provides further evidence for the widespread 
presence of PAHs in QSOs.
  
Similar PAH/FIR and [NeII]/FIR ratios are found
in QSOs and in starburst-dominated ULIRGs and lower luminosity
starbursts.  We conclude that the typical QSO in our sample has
at least 30\% but likely most of the far-infrared 
luminosity ($\sim 10^{10\ldots 12}\Lsun$) arising 
from star formation, with a tendency for larger star formation contribution at
the largest FIR luminosities.

In the QSO sample, we find correlations
between most of the quantities studied including combinations of AGN tracers
and starburst tracers. The common scaling of AGN and starburst 
luminosities (and fluxes) is evidence for a starburst-AGN connection in 
luminous AGN. 
Strong correlations of far-infrared continuum and starburst related quantities
(PAH, low excitation [NeII]) offer additional support for the starburst 
origin of far-infrared emission. 

\end{abstract}


\section{Introduction}
The infrared properties of luminous active galactic nuclei 
(AGN) and ultraluminous infrared galaxies (ULIRGs) hold clues 
for the understanding of galaxy formation, the star formation history of 
the universe and the connection between black hole and galaxy formation 
and evolution. Physical insight into the co-evolution of AGN and 
star formation can be gained by study of the low redshift members of 
these populations. The Infrared Astronomical Satellite (\IRAS) 
and the Infrared Space Observatory (\ISO) have allowed the first comprehensive
far-infrared (FIR) studies of extragalactic 
sources and some high quality mid-infrared (MIR) observations on the brighter 
sources \citep[see e.g. the review of ][]{genzel00}. Issues discussed 
in \IRAS\ and \ISO\ studies include the nature of ultraluminous 
infrared galaxies and their starburst and AGN energy sources
\citep{genzel98,klaas01}, and the SEDs of QSOs and their evolutionary implications
\citep{haas03}. A  2.5 to 45 $\mu$m
spectral inventory of starburst and AGN prototypes \citep{sturm00} and 
detailed mid-infrared spectroscopic investigations of local lower luminosity 
AGN \citep[][hereafter S02]{clavel00,sturm02} have addressed AGN unification 
and the starburst-AGN connection.  

The {\it Spitzer Space Telescope} (hereafter \spitzer ) has enabled us to 
build upon previous work with significantly improved sensitivity.
This work is part of a series of papers describing the results of  
the Quasar and Ulirg Evolution STudy (QUEST), focussing on mid-infrared
spectroscopy of a sample of 54 QSOs and 
ULIRGs at redshifts z$\lesssim$0.3 with the Infrared Spectrograph 
\IRS\ onboard \spitzer .  The
sample was selected with the aim of investigating the possible 
connections between those two groups of luminous 
(L$_{Bol}\gtrsim 10^{12}\Lsun$)
objects in the nearby universe, and is closely connected to optical and 
near-infrared studies of the morphology and dynamical properties of these
two populations \citep[e.g.][]{dasyra06a,dasyra06b,veilleux06}. 
The purpose of the present paper is 
to study starburst signatures in QSO host galaxies and to show that most, 
and in some cases all, of the far-infrared luminosity in QSOs 
showing strong PAH 
features is due to starburst activity. A forthcoming complementary paper 
addresses the spectral energy distribution (SED) of QSOs including the 
clear far-infrared 
starburst contribution to the $\lambda > 30 \mu m$ continuum of QSOs.

To investigate the link between AGN activity and star formation and the extent
to which they occur simultaneously, it is important to quantify the star 
formation activity in QSO hosts.  Such measurements are made difficult, 
however, by the observational problems of detecting
star formation tracers in the presence of extremely powerful AGN emission.
SED studies based on the \IRAS\ and \ISO\ space missions have established QSOs 
as sources of (sometimes) strong far-infrared emission.
\citep[e.g.][]{neugebauer86,haas03}. In addition to a nonthermal continuum
that is detectable in the infrared only in flat spectrum radio-loud QSOs
\citep[e.g.][]{haas98}, a
strong far-infrared emission component is often observed, at varying levels 
with respect to the strong AGN mid-infrared continuum.
Due to its steep falloff in the submillimeter regime, the origin of
this far-infrared emission must be
thermal emission of optically thin dust \citep{chini89,hughes93}. 
While the warmer
T$\sim$200K dust, which dominates the mid-infrared SEDs of QSOs, is 
generally accepted to be predominantly AGN heated,
there is still considerable dispute about the origin of the cooler T$\sim$50K
emission often dominating the far-infrared. Direct heating by the powerful 
AGN, but at distances ensuring sufficently low temperatures, is one
possibility \citep[e.g.][]{sanders89,haas03,ho05}. Other models prefer
an origin in vigorous star formation in the QSO host \citep[e.g.][]{roro95}.
\cite{roro95} used radiative transfer modelling to infer an SED of AGN heated 
dust that, in $\nu$L$_\nu$ units, peaks in the mid-IR and decays towards the 
far-infrared, a feature shared by many other such models. In the QSO SEDs 
that are often flat over a wide wavelength range including the far-infrared, 
the far-infrared component is then plausibly ascribed to a 
component with an SED similar to that of a star-forming galaxy, in accordance
with evidence for coexistence of star formation and AGN in spatially resolved
lower luminosity AGN. \Citet{roro95} found a tight correlation of AGN 
optical emission and mid-infrared continuum and a weaker correlation 
between optical and far-infrared emission, which is supporting the view 
that the far-infrared does not result directly from AGN heating but that 
there is a connection between AGN and starburst luminosities in the QSOs.  

Our goals are to quantify star formation in QSO hosts and to estimate its 
contribution to the
the far-infrared emission. In the mid-infrared, the contrast between
the emission from possibly dust-obscured star formation and from the
central AGN is favourable,
and established star formation tracers are available. We use two such tracers:

(1) The mid-infrared broad aromatic `PAH' emission features arise in regions 
of the interstellar medium of a galaxy where their aromatic carriers are 
present, and where their transient excitation is made possible by a 
non-ionizing ($<13.6$eV) soft UV radiation field. This is the case in the 
photodissociation regions (PDRs) that accompany Galactic star formation 
regions \citep[e.g.][]{verstraete96}, as well as in the diffuse interstellar
medium where they are excited by the general interstellar UV radiation field
that has leaked from its OB star origins to large scales 
\citep[e.g.][]{mattila96}.  PAHs have
been used as a quantitative tracer of star formation in galaxies
\citep[e.g.][]{genzel98, foerster04, calzetti05}. 
Metallicity above 0.2 solar is
 a prerequisite for strong PAH emission \citep{engelbracht05},
a condition that is probably safely met for local QSO hosts. Destruction of 
the PAH carriers in energetic environments but survival in starburst PDRs 
(though not in \hii\ regions proper) is 
key for the use of PAH features as diagnostic. The PAH features 
are absent from the hard radiation environment of AGNs according to both
empirical \citep[e.g.][]{roche91,lefloch01,siebenmorgen04a} and theoretical 
\citep{voit92} studies. The latter suggest that PAH molecules hit by single
energetic EUV/X-ray photons can be efficiently destroyed by photo-thermo 
dissociation or Coulomb explosion. Since AGN are copious emitters of hard 
photons, PAH molecules near AGN will be destroyed unless shielded by a large
obscuring column. 

(2) The low excitation fine-structure emission lines like \ne212m \
are among the dominant
emission lines of HII regions. Observations of starburst galaxies, as 
well as a combination of evolutionary synthesis and photoionization 
modelling, show \ne212m \ to be stronger than higher excitation mid-infrared 
lines ([NeV], [OIV], [NeIII]) in typical ionized regions excited by 
young stellar populations \citep{thornley00,verma03}. Use of low excitation 
lines as star formation tracers requires, however, the consideration of 
possible contributions from the AGN Narrow Line
Region (NLR) which can be significant despite the generally higher excitation 
of such regions \citep[e.g.][]{spinoglio92,alexander99} compared to starburst 
\hii\ regions.

Section 2 of the paper describes the sample, observations and data 
reduction used to obtain the line and continuum fluxes
in our sources. Emission lines that are relevant to the present study 
are tabulated for all sources.
In Section 3 we discuss the widespread presence of PAH emission and its 
relation to other components of the QSO spectra.
Finally, Section 4 addresses the issue of star formation in host galaxies of 
QSOs, shows the importance of this process and compares our results with 
earlier findings. In a forthcoming paper,
we discuss the implications of our results for QSO SEDs in general.
We adopt $\Omega_m =0.3$, $\Omega_\Lambda =0.7$ and
$H_0=70$ km\,s$^{-1}$\,Mpc$^{-1}$.

\section{The QUEST PG QSO sample: Observations and reduction}

\subsection{The sample}

As part of the \spitzer\ spectroscopy project QUEST (PID 3187, PI Veilleux)
we are studying QSOs, ultraluminous infrared galaxies, and the possible
evolutionary connection between the two using the infrared spectrograph 
\IRS\ \citep{houck04}. The QSO sample is largely drawn from that of 
\citet{guyon02} and \citet{guyon06}. It consists of
Palomar-Green (PG) QSOs \citep{schmidt83} and
covers the full ranges of bolometric luminosity 
\citep[$\sim 10^{11.5-13}$\Lsun\ based on the B band absolute magnitude and 
the SED of][]{elvis94}, radio
loudness, and infrared excess (log(\nulnu60 /L$_{Bol}$)$\sim$ 0.02--0.35) 
spanned by the local members of the PG QSO sample  \citep[see also][for a 
recent view on selection effects in the PG sample]{jester05}. 
B2 2201+31A is not a PG QSO but is
in the sample because its B magnitude actually satisfies the PG QSO 
completeness criterion of \citet{schmidt83}. It is one of the five radio-loud 
systems in our sample.
At the sample's maximum redshift of 0.325, important emission lines like 
[O\,IV] 25.89$\mu$m
stay within the IRS spectral range for all objects. The QUEST sample used in 
this paper includes 23 of 32 objects from the Guyon sample. We exclude here
two recently observed QUEST objects that are not yet fully processed by the 
\spitzer\ pipeline. We add two Palomar-Green objects from the Guyon et al. 
sample previously observed 
by \spitzer\ (PG0050+124 = IZw1; \citet{weedman05} and PG0157+001 = Mrk1014; 
\citet{armus04}) and two PG QSOs from another project
(PID 20241, PI Lutz). Table~\ref{tab:targets} lists names and redshifts of 
all 27 QSOs in our sample. In total, our sample covers a range from 
M$_B$ -21 to -26, with median -23.3. In the remainder of this paper, we 
will compare
some aspects of the PG QSOs to ultraluminous infrared galaxies whose 
properties will 
be presented in more detail in an upcoming paper based on QUEST data 
(Veilleux et al., in preparation).

\subsection{Data reduction and line and continuum measurements}
For the QUEST sample, spectra were taken both at 5-14$\mu$m in the low 
resolution (SL short-low) mode and at 10--37$\mu$m in the high resolution 
(SH short-high and LH long-high) modes of the \IRS\ . Slit widths of 
3.6\arcsec\ to 
11.1\arcsec\ include much of the QSO hosts as well as the vicinity of the AGN. 
Our data reduction starts with the two-dimensional basic calibrated data
(BCD) products provided by version 12 of the \spitzer\ pipeline
reduction. We used our own IDL-based tools for removing outlying values 
for individual pixels and for sky subtraction
and SMART \citep{higdon04} for extraction of the final spectra. Small 
multiplicative corrections were applied to stitch together
the individual orders of the
low resolution and high resolution spectra, as well as additive corrections
for residual offsets still found between the low resolution spectra
and the SH and LH high resolution spectra after zodiacal light correction 
of the latter.
Emission line fluxes were measured using fits of Gaussian lines superposed
on a local continuum.
PAH fluxes were mostly measured by simultaneously fitting Lorentzians to the
main 6.2, 7.7, and 8.6$\mu$m features superposed on a 5.3 to 9.6$\mu$m 
(rest frame) continuum approximated by a second order polynomial, and in a few
cases by combined Lorentzian and continum fits over smaller ranges.  We use 
below the flux for the brightest (7.7$\mu$m) feature as PAH strength 
indicator F$_{\rm PAH}$. In PG0050+124, where this feature
was present but difficult to quantify in the presence of strong 
silicate emission,
we have estimated the flux scaling from the flux of the well-detected 
6.2$\mu$m PAH feature, using the 7.7/6.2 feature ratio measured in 
the starburst/AGN NGC\,6240 
\citep{lutz03,armus06}. Flux upper limits (3$\sigma$) were derived 
adopting typical widths for the lines ($\sim$600km/s) and broad features
($\sim 0.6\mu$m for the 7.7$\mu$m feature).
In one source (PG1307+085) we could only analyse the high resolution 
spectra since low resolution data are proprietary to another program.
This limits our sample to 26 objects for analysis related to PAH emission.
QUEST data for ultraluminous infrared galaxies that are used for comparison
with the QSOs were processed in the same way.

\begin{figure}
\plotone{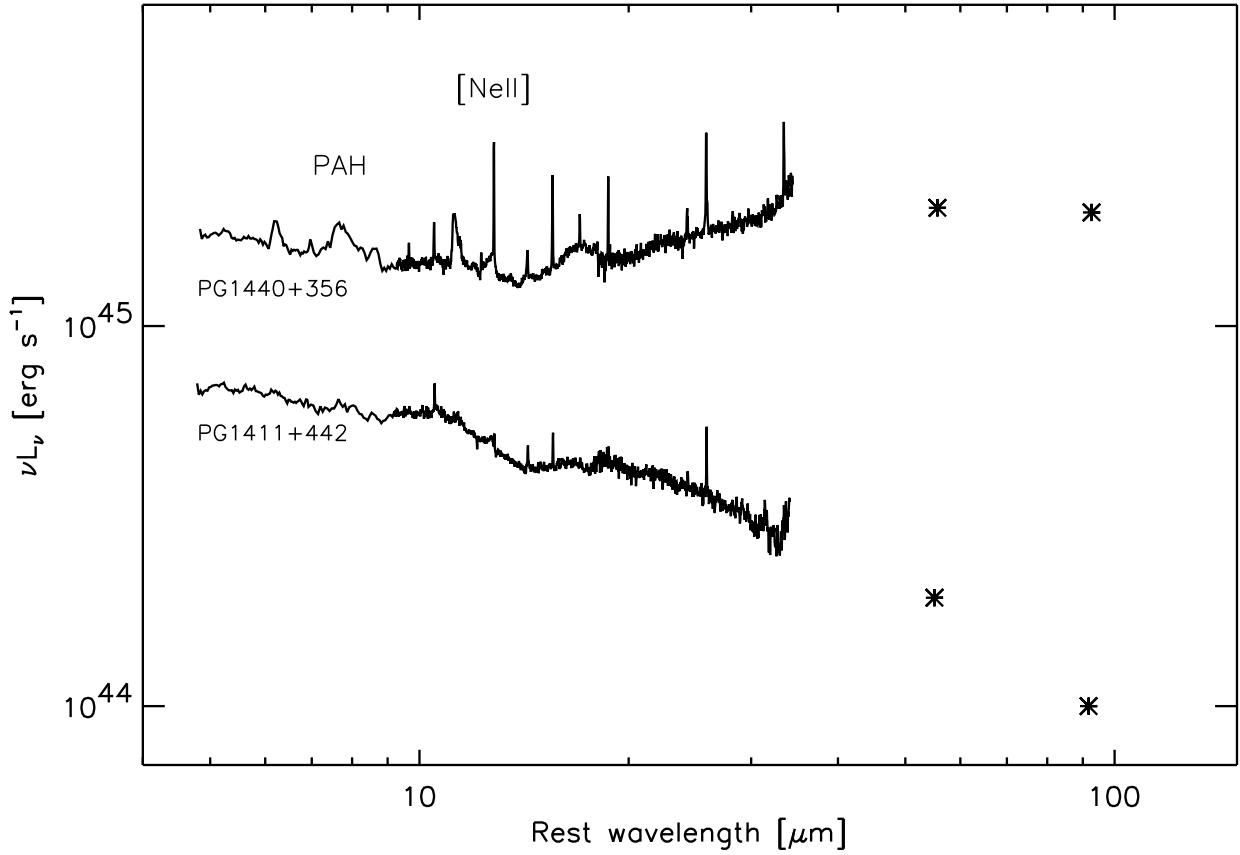}
\caption{IRS spectra of two PG QSOs, combined with far-infrared
photometric fluxes. These two nearby sources were selected to illustrate
the full range between sources with strong PAH, [NeII], and far-infrared
emission and others with weak emission in all these tracers. The spectrum
of PG1440+356 has been multiplied by a factor 4 for clarity.}
\label{fig:twospec}
\end{figure}

Figure~\ref{fig:twospec} shows individual example spectra of two PG QSOs 
illustrating the broad differences between QSOs with strong and weak 
PAH emission. 
Table~\ref{tab:targets} also lists the continuum flux densities in several 
mid-  and far-infrared bands. In the mid-infrared, the quality of the \IRS\ 
spectra is superior to most pre-existing \IRAS\ or \ISO\ photometry. For 
that reason and to characterise the continuum at wavelengths reasonably free of
emission features, we have derived average observed flux densities over narrow
bands ($\Delta\lambda /\lambda\sim 0.07$) centered at several rest wavelengths.
The shortest wavelength continuum point at 6$\mu$m is shortward of the main 
PAH complex; the 15$\mu$m point is between the two
silicate emission peaks at about 10 and 18$\mu$m 
\citep{siebenmorgen05,hao05,sturm05}, but still 
partly affected by silicate emission if
present. The longest wavelength point at 30$\mu$m is near the long end 
of the \IRS\ spectra and already 
beyond the strongest part of the longer wavelength silicate feature.   
We later use continuum luminosities at these rest wavelengths: 
6$\mu$m, 15$\mu$m, and 30$\mu$m, defined as $\nu L_{\nu}$. 

Far-infrared fluxes have been taken
from the literature, usually giving preference to \ISO\ based results over
\IRAS\ based ones, because the smaller effective \ISO\ beams reduced the 
susceptibility to cirrus contamination at 100$\mu$m.
The far IR luminosity, L(FIR), is often obtained from 
${\rm F_{FIR} }$  which is defined as
${\rm F_{FIR}=1.26 (S_{100}+2.58 \times S_{60}) \times 10^{-18} W/cm^2 }$,
where  ${\rm S_{100} } $ and ${\rm S_{60}}$ are flux densities in Jy
\citep[e.g.][]{sanders96}. In order to use a consistent definition over 
the full z$\leq$0.33 redshift range of our sample and to reduce the 
sensitivity to galactic cirrus contamination that is most problematic  at
100$\mu$m, we adopt a far-infrared luminosity $\nu L_\nu$ which is solely based
on the flux at rest wavelength 60$\mu$m interpolated from the photometry
in the literature. Even nearby QSOs are typically close to the 
far-infrared detection limits of the previous space missions.
In six cases where only one of S$_{100}$ or S$_{60}$ was detected we 
estimated the other flux as the lower of the measured limit and an 
extrapolation using the detected flux and the ratio 
S$_{60}$/S$_{100}$ = 0.93 
that is the median for the part of the sample detected in both bands.  
The flux at 60$\mu$m rest wavelength was interpolated linearly
between the observed 60 and 100$\mu$m points.
For seven PG QSOs in our sample that are undetected in both bands,
upper limits on  $\nu L_\nu$ at 
60$\mu$m rest wavelength were derived in an analogous way interpolating
linearly between the limits at observed wavelengths of 60 and 100$\mu$m.
Since QSO fluxes are close to the detection limits of the original \IRAS\
and \ISO\ references and because of residual cirrus contamination, 
uncertainties of order 30-50\% for the far-infrared fluxes may well occur.

\section{Results}

Table \ref{tab:emission_lines} lists the intensities of the three emission
lines used in the present analysis, \ne212m , \nevIR , \oivIR\ and of 
the emission feature \pahIR. The table includes detections 
as well as upper limits. A more comprehensive list including more lines
will be given in a forthcoming paper.

\subsection{PAH emission is common in QSOs}

As is clear from inspection of Table \ref{tab:emission_lines}, some of 
the QSOs in our sample show only high excitation lines, likely 
originating in the 
NLRs of these objects. Others 
(11 of 26) show  prominent \pahIR\ and other PAH features, indicative of 
significant star formation. The existence 
of ``composite'' sources (see S02 for definition and references
to ISO-based studies) that show both AGN and starburst properties, has been 
known for several years. Such sources contain both powerful 
star forming
regions, giving rise to the observed PAH emission through near-UV heating, 
and high luminosity NLRs that are
excited by the central radiation source. Since the
star forming activity is a property of the host galaxy, and 
the NLR line luminosities scale with the AGN continuum
luminosity, the intensity of the starburst 
emission features relative to the continuum might be expected to be
weaker in AGN whose luminosities are in the QSO regime. Because the S02 sample 
included only three objects with QSO-like luminosities (i.e.
L(FIR)$>5 \times 10^{11}$\Lsun) and suffered from the limited S/N 
achievable with \ISO , a quantitative verification of this idea was 
not possible up to now.

Our new sample includes 27 high luminosity AGN and it is
therefore important that the fraction of clear PAH emitters in the sample 
is large. This is strengthened further by the detection of
PAH 7.7$\mu$m and 11.3$\mu$m peaks in the average spectrum of 
the 15 QSOs not showing 
{\em individual} PAH detections (Fig.~\ref{fig:plotpgpahnonpah}, bottom). 
Since this average spectrum excludes sources with individual PAH 
detections, the PAH emission seen is unlikely 
to be due to a few PAH-strong sources only, in an otherwise largely PAH-free
group. It also has the implication that typically the true PAH fluxes of the
PAH nondetections cannot be far below our limits.
This high incidence of PAH emission is our first major conclusion 
and is further discussed in the following sections.
 
\subsection{Trends with level of PAH emission}
\begin{figure}
\epsscale{0.8}
\plotone{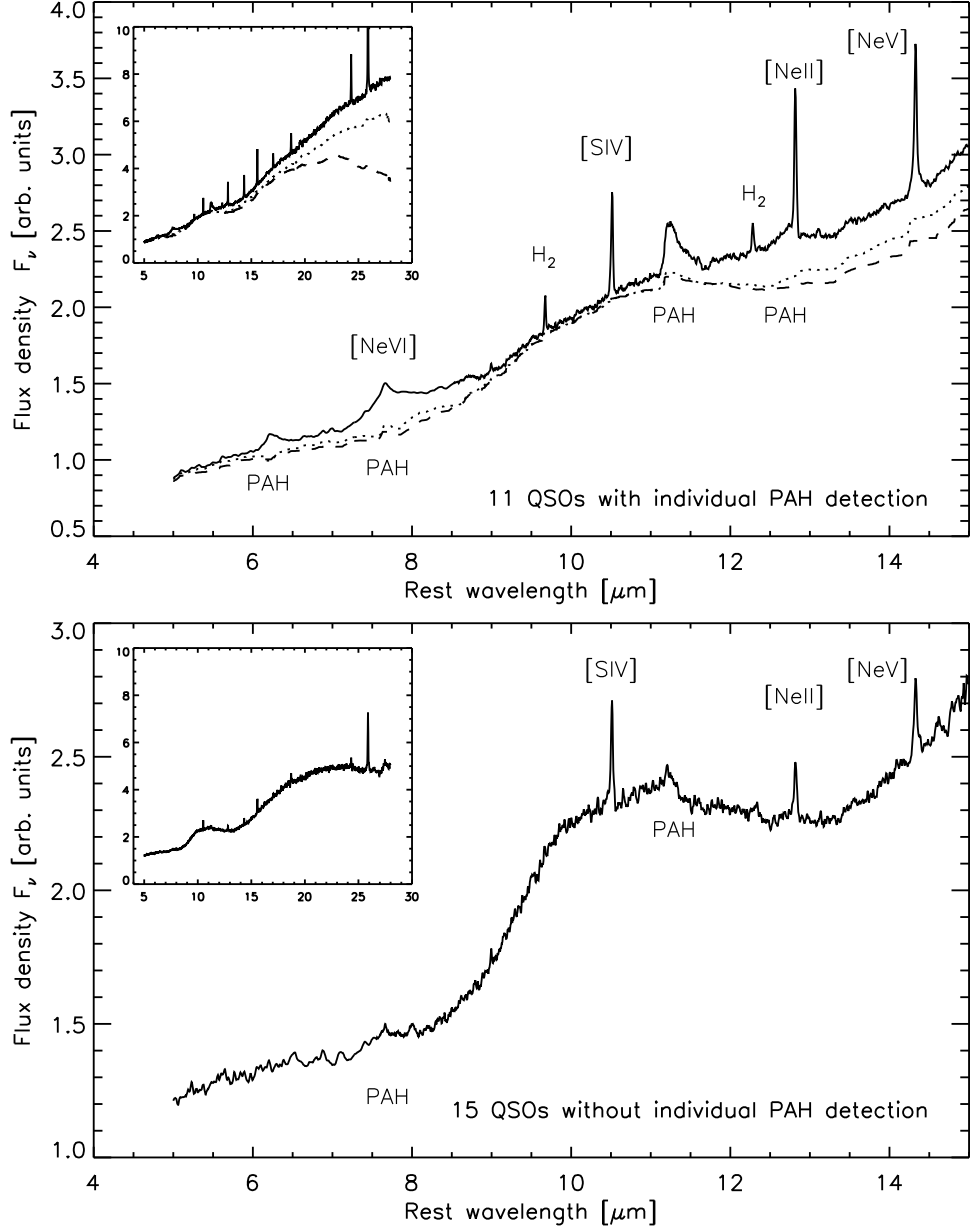}
\caption{Average spectra of QSOs. Individual spectra have been normalized
to the same total 5-25$\mu$m flux before averaging. Note the change in 
spectral 
resolution from low at short wavelengths to high at $\gtrsim 9\mu$m rest
wavelength. Top: 11 QSOs for which 
\pahIR\ is detected in the individual spectra. The dotted line is the same
spectrum after subtracting an M82 spectrum scaled to the PAH features, and 
the dashed line after subtracting a starburst-dominated ULIRG spectrum scaled
to the PAH features. Emission
line residuals have been removed for clarity. The inset repeats the same 
spectra with different plot scaling, emphasizing the wider range SED
trends. Bottom: 15 QSOs for which
\pahIR\ is not detected individually. A broad maximum near 7.7$\mu$m rest 
wavelength as well as an 11.3$\mu$m feature is detected in the average, 
however, indicating a high incidence
of PAH emission in the contributing spectra.}
\label{fig:plotpgpahnonpah}
\end{figure}

Before embarking on an analysis of the correlations of different starburst and
AGN tracers, we use the two average spectra of the PG QSOs that are 
individually detected in PAH emission (11 objects) or individually 
undetected in PAH emission (15 objects) 
(Fig.~\ref{fig:plotpgpahnonpah}) to identify some of the salient trends
in our data.
The two spectra have been obtained by averaging the individual spectra after
normalizing to the same total flux in the mid-infrared 5-25$\mu$m rest 
wavelength region. We caution that significant variations are present
also within those two groups.

(i) \pahIR\ is 2.5 times stronger (relative to the total 
5-25$\mu$m mid-infrared flux) in the average of the
objects with individual PAH detections. This is a natural consequence of 
the object grouping for the two average spectra. As noted, the relatively
small difference argues for the presence of PAH emission in most of the objects
that did not have individual detections.

(ii) The far-infrared continuum emission is relatively stronger 
compared to the mid-infrared continuum
in the objects with PAH detections. This is apparent in several ways.
Firstly, all PAH detections are also 
far-infrared (60$\mu$m rest-frame) detections while 7 of 15 PAH 
nondetections are also not detected in the far infrared. Secondly, 
the mean ratio
$\nu F_\nu (60\mu m) / \nu F_\nu (6\mu$m) is 1.72$\pm$0.54 for the 11 PAH 
detections,
1.00$\pm$0.15 for the 8 FIR-detected PAH nondetections, $<0.88$ for the
7 objects undetected in both far-infrared and PAH, and finally $<$0.94 
for all 8+7 PAH nondetections. The values for the two upper limits 
assume that the far-infrared fluxes are less than or
equal to the measured limits for the individual far-infrared nondetections. 
Thirdly, inspection of the average
\IRS\ spectra in Fig.~\ref{fig:plotpgpahnonpah} shows significant 
differences in the extrapolation to 60$\mu$m: The average spectrum
of the PAH detections continues to rise beyond 25$\mu$m, while there is
an indication for a downturn in the average of the PAH nondetections beyond
this wavelength. Regardless of uncertainties in the extrapolation 
that are related to the assumed intrinsic spectral shape and to
technical issues like \IRS\ spectral response calibration and zodiacal 
light subtraction, 
the 60$\mu$m to 6$\mu$m flux density ratio must be significantly larger
in the average spectrum of the sources with detected PAHs. We 
estimate that this ratio is larger for the PAH detections by a factor 
2 to 3.  

We have also tentatively grouped the PAH nondetections into two average 
spectra for the 8 far-infrared-detected and 7 far-infrared-nondetected 
sources. 
While the statistics and S/N are poorer in these samples, there are 
indications that the sources undetected in both 
tracers are at the end of the physical trends described in (i) and (ii) -- 
little PAH, weak far-infrared, and indication for more of a downturn at 
$\lambda >25\mu$m; that is, they are not just fainter overall.

(iii) The lower excitation [NeII] line is stronger in the QSOs detected
in PAHs. Normalized by the higher excitation lines [SIV] and [OIV], the
[NeII] line is 1.9 and 1.7 times stronger in the QSOs detected in PAHs 
compared to those undetected in PAHs.

(iv) The broad 10$\mu$m silicate emission peak is apparently weaker in 
the sources with detected PAHs. We argue this is largely an artefact of having 
a stronger starburst component with its PAH emission. This is 
illustrated by the dotted line in the upper panel
of Fig.~\ref{fig:plotpgpahnonpah}, which shows the average spectrum after 
subtracting a PAH-dominated M82 spectrum \citep{sturm00} roughly scaled 
to the PAH features, and the dashed line, which shows the equivalent result
using the average of 12 starburst-dominated QUEST ULIRGs (see \S 4.1) as 
a starburst template.
As noted by, e.g., \citet{sturm00}, a main difference between different 
starburst templates is in the level of increase of the 
$\lambda\gtrsim 12\mu$m very small 
grain continuum, even for similar shorter wavelength PAH spectra. Our two 
choices illustrate the effect of such a template variation. 
The two PAH complexes around  7 and 12$\mu$m and the far-infrared upturn
serve to `fill in' the minima between and around the two silicate emission 
peaks. This is analogous to the well-known difficulty of quantifying silicate
{\em absorption} in spectra with strong PAH emission. We note that subtracting 
a starburst
spectrum from the PAH-strong average QSO spectrum, in addition to recovering
the correct pronounced silicate emission, also indicates a flatter or
decreasing extrapolation to beyond 25$\mu$m, similar to what is 
seen in the PAH-weak average QSO spectrum.

(v) Molecular hydrogen emission in the S(1), S(2) and S(3) rotational
lines has a larger equivalent width in the PAH-strong average spectrum, by 
factors 3-5 comparing these detections with the detections/limits for the
PAH-weak average spectrum.  

All these trends are consistent with a starburst component (containing strong 
PAH and far-infrared continuum, low excitation fine-structure line, 
and possibly molecular hydrogen emission) being superimposed in increasing 
proportion on a pure AGN spectrum (consisting of warm and hot 
dust continuum, silicate, and high excitation line emission).
 
\section{Discussion}

\subsection{Nature of the QSO far-infrared emission and the starburst-AGN 
connection}

We use our sample to compare several AGN- and starburst
related quantities observed in the PG QSOs, in an attempt to identify the
likely origin of the far-infrared emission. We compare the starburst tracers
in these QSOs with those of pure starbursts from two samples: 
(i) A subset of starburst-dominated QUEST ULIRGs observed with \IRS\  
(`SB-ULIRGS'). In
order to restrict ourselves to the ULIRGs with the highest starburst 
contribution to their infrared luminosity, we 
require these objects to have no \oivIR\ detection, a peak ratio of the 
\pahIR\ feature to its local continuum of at least 1, and a 5-10$\mu$m spectrum
in which visual inspection shows that the absorption features like the 
6$\mu$m ice feature are not dominant,
although sometimes present. The last criterion is used to avoid ambiguities
concerning the internal energy sources of the most heavily obscured ULIRGs 
\citep{spoon04a,spoon04b}. 
(ii) A small sample of six \ISO -observed local 
starbursts (M82, NGC253, NGC1808, IC342, NGC3256, NGC7552) for which 
the PAH emission is measured without significant
aperture corrections relative to the far-infrared data from \IRAS , \ISO\
or the Kuiper Airborne Observatory \KAO.
The ratio of PAH and far-infrared emission is 
known to vary somewhat with physical conditions \citep[see e.g. discussion 
in \S 4.2 of][]{lutz03}, decreasing with the average intensity of the 
radiation field \citep[e.g.,][]{dale01}. Because of these trends with ISM 
conditions, we add as another group of comparison objects (iii) twelve 
`FIR quiescent' galaxies (with particularly low L$_{FIR}$/L$_{B}$ and 
low S$_{60}$/S$_{100}$) 
from the sample of normal galaxies of \citet{lu03}, for which we convert their
PAH fluxes measured in ISOPHOT-S spectra to our measurement procedure and 
apply aperture corrections based on their PAH-dominated ISOCAM LW2 images.
We add to this group NGC 891 that has been mapped with ISOPHOT-S by 
\citet{mattila99}, arriving at a sample of 13 FIR-quiescent objects. 
The three comparison samples cover a range in luminosity as well as in the 
average intensity of their interstellar radiation fields that is reflected 
in their mid-to far-infrared SEDs.

\begin{figure}
\epsscale{1.0}
\plotone{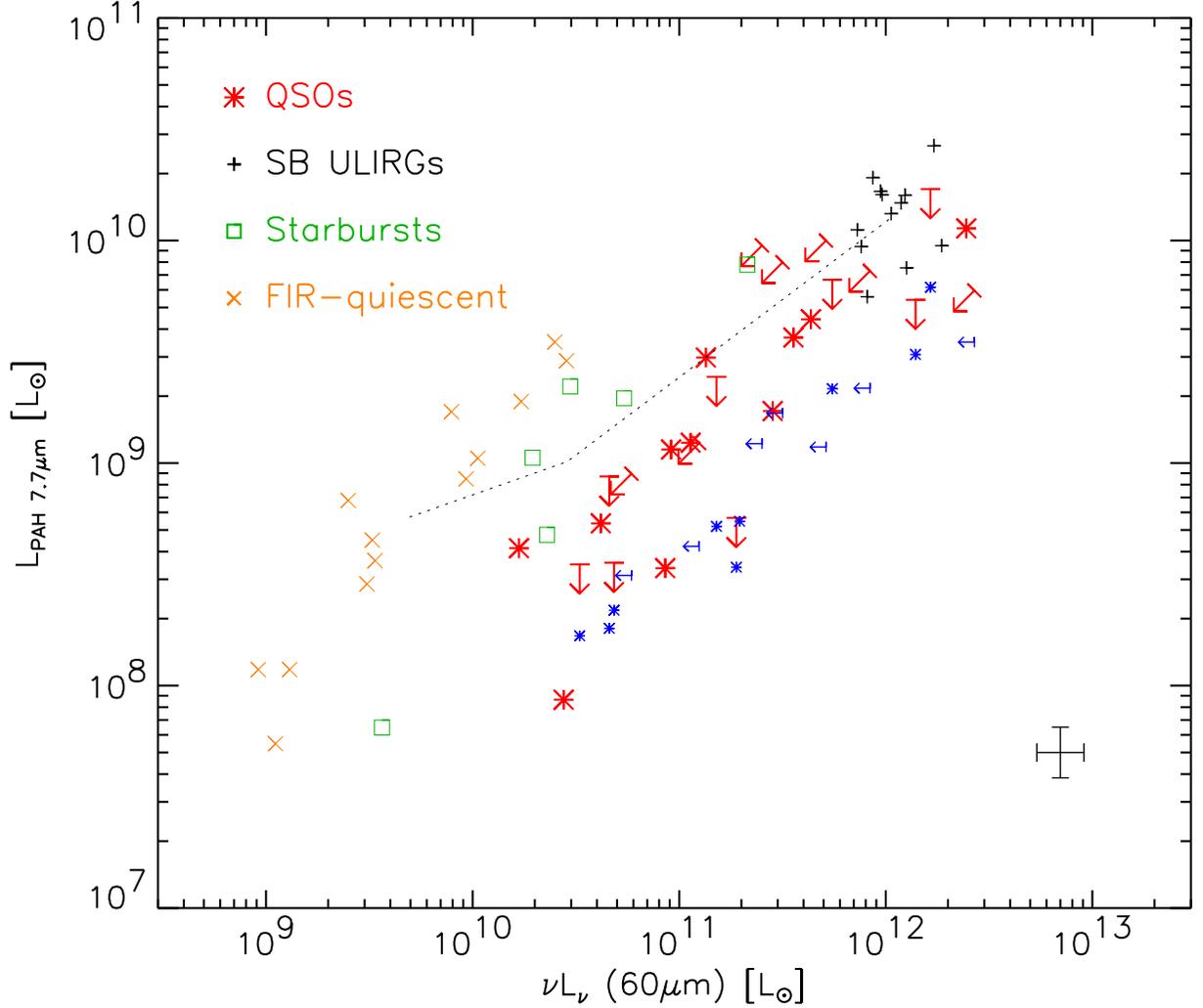}
\caption
{Luminosities L(\pahIR)\ vs. L(60$\mu$m) for the QSOs.  Starburst dominated 
ULIRGs, lower luminosity starbursts, and FIR-quiescent galaxiess are added for 
comparison. The thin dotted line connects the mean locations of these three 
groups of comparison objects.
The small blue symbols repeat the QSOs without individual PAH detections,
but assuming that the individual ratios of \pahIR\ to rest 
frame 5-25$\mu$m flux
are the same as in the average spectrum of Fig.~\ref{fig:plotpgpahnonpah}
(bottom). A 30\% 1$\sigma$ uncertainty is indicated in the lower right corner.}
\label{fig:pah_fir}
\end{figure}

\begin{figure}
\epsscale{0.8}
\plotone{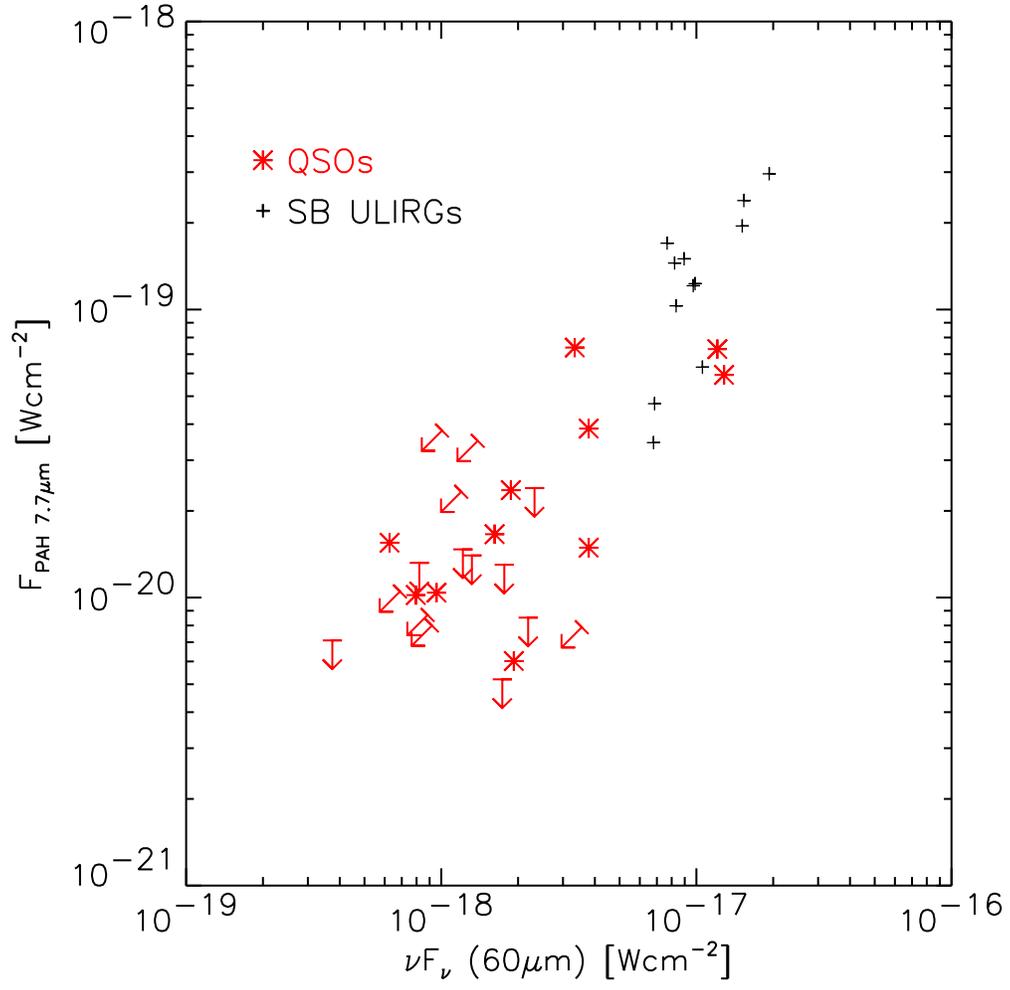}
\caption{Fluxes F(\pahIR)\ vs. F(60$\mu$m) for the QSOs and for starburst 
dominated ULIRGs.}
\label{fig:pah_fir_flux}
\end{figure}

\begin{figure}
\plotone{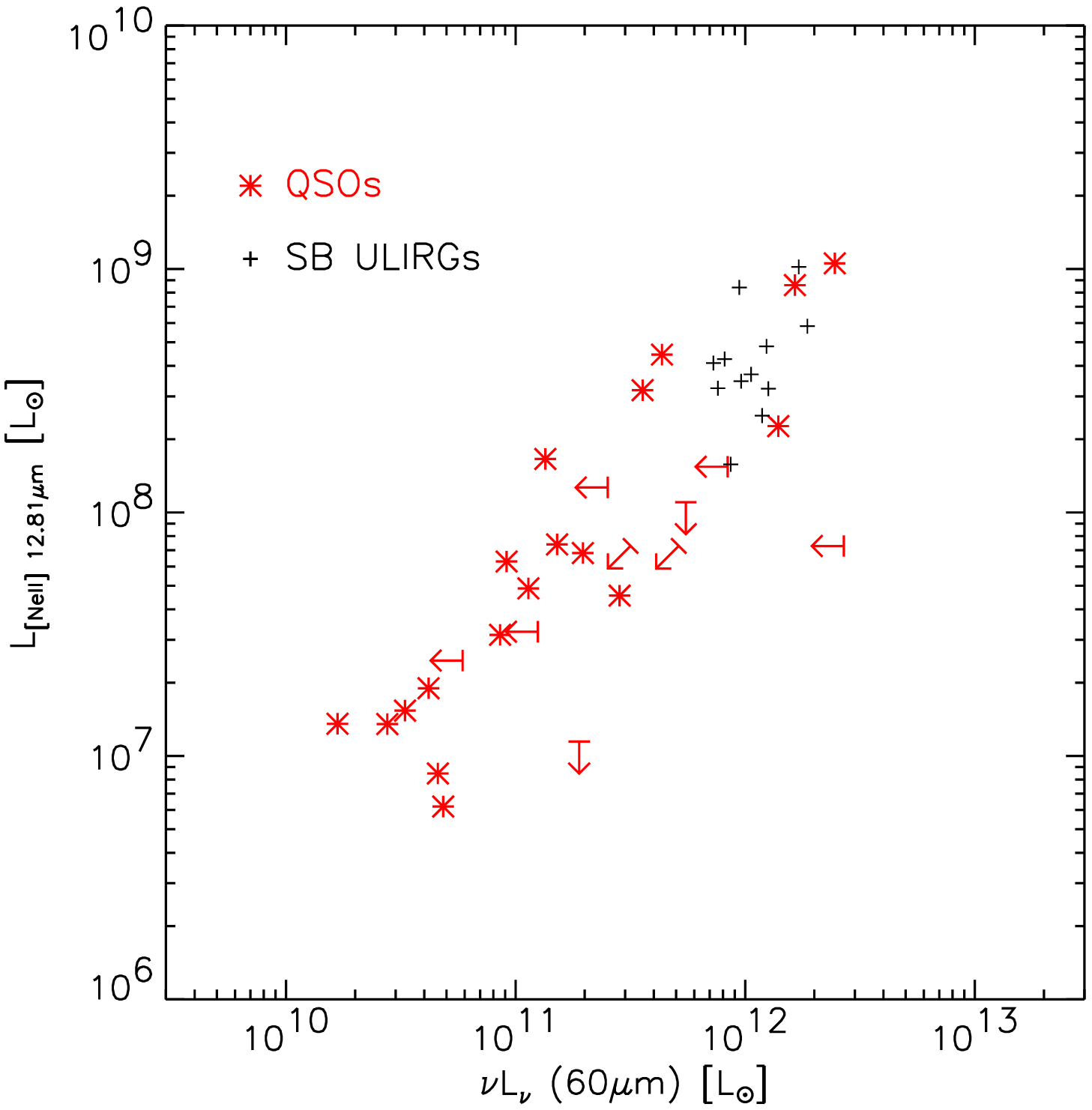}
\caption
{Luminosities L(\ne212m ) vs. L(60$\mu$m) for the QSOs and for 
starburst dominated ULIRGs.}
\label{fig:ne2_fir}
\end{figure}

An important new result is the clear correlation between L(\pahIR) and 
L(FIR) for the PAH-containing QSOs. 
Figs. \ref{fig:pah_fir} and \ref{fig:pah_fir_flux} compare the 
luminosities and fluxes of \pahIR\ and 60$\mu$m continuum. 
The QSOs with PAH detections
and the starburst-dominated ULIRGs follow the same trend.
Fig. \ref{fig:ne2_fir} shows the equivalent continuous trend
from QSOs to starburst-dominated ULIRGs for the comparison of \ne212m \ and 
60$\mu$m continuum.

These trends also imply almost identical mean ratios 
for the QSO and starburst-dominated ULIRG 
populations. For the 11 QSOs with both PAH and far-infrared emission
detected we find ${\rm <L(PAH) / L(FIR)>=0.0110 \pm 0.0021}$ 
while for the 12 starburst-dominated ULIRGs we get 
${\rm <L(PAH)/L(FIR)>=0.0130 \pm 0.0015}$. Similarly, comparing 
\ne212m \ and far-infrared emission we get for the 18 QSOs with both 
quantities detected 
$\rm <L([NeII]) / L(FIR)>=(5.15 \pm 0.73)\times 10^{-4}$ 
and for the 12 starburst-dominated ULIRGs 
$\rm <L([NeII]) / L(FIR)>=(4.22 \pm 0.57)\times 10^{-4} $.
In both cases, the ratio of the star formation indicator and 
far-infrared emission
is the same in QSOs and in starburst-dominated ULIRGs -- the observations
are consistent with starbursts producing all of the QSO 
far-infrared emission if their specific properties are similar to those in
starburst dominated ULIRGs. 

This argument depends on the adopted specific properties of the comparison
starforming galaxies, because of the changes of the PAH to 60$\mu$m
ratio with ISM properties mentioned above. For ULIRG starbursts we 
obtained ${\rm <L(PAH)/L(FIR)>=0.0130 \pm 0.0015}$. For the six 
lower luminosity 
starbursts for which the ratio of PAH and far-infrared can be derived
from \ISO\ data without significant aperture corrections, we
obtain \\ ${\rm <L(PAH)/L(FIR)>=0.0399 \pm 0.0087}$. Finally, for the sample of
13 FIR-quiescent normal galaxies we get ${\rm <L(PAH)/L(FIR)>=0.126 \pm 0.016}$
with individual objects like NGC 891 reaching up to a value 0.2.

Comparing these three ratios to the average of the QSOs, we find that almost 
all, at least $\sim$30\%, and at least $\sim$10\% of the QSO far-infrared 
emission would be due to  non-AGN sources (that is current and recent star 
formation), adopting the values for ULIRGs, starbursts and FIR-quiescent 
galaxies as templates. Irrespective of the adopted template, PAH-based values 
are lower limits since for QSOs there is the interesting 
possibility of additional weakening of PAH emission in star-forming 
regions that are also exposed to AGN radiation, by destruction of the 
PAH molecules. This may be particularly relevant 
if some of the far-infrared emission is generated by star formation occurring
in compact clusters or disks very close to the AGN \citep{davies04a,davies04b}.
This makes values towards the upper end of the range for the non-AGN
contribution more likely. 

Using only PAH and far-infrared evidence, a formal solution is possible which 
maximizes the AGN fraction of the far-infrared emission by assigning the PAH 
emission to a very FIR-quiescent galaxy. The QSO PAH luminosities are 
on average larger than those of the FIR quiescent galaxies, but the most 
luminous 
quiescent objects are of a similar PAH luminosity as the typical QSOs 
(Fig.~\ref{fig:pah_fir}). Comparing these objects to QSOs, the AGN would
then contribute roughly 90\% of the QSO far-infrared flux. These objects yield 
biased comparisons, however, since they are specifically selected for FIR
quietness. A fairer comparison would need to populate the PAH/FIR plane of 
Fig.~\ref{fig:pah_fir} with a large and complete sample of non-AGN galaxies.
While such a sample is currently not available, we qualitatively indicate in 
Fig.~\ref{fig:pah_fir} the expected locus of such non-AGN galaxies by 
connecting the locations of the averages of our three small comparison 
samples (dotted line). While the FIR quiescent
objects are likely more PAH luminous than the average galaxy of the same FIR 
luminosity, and thus their point placed too high, the overall trend of PAH 
flux with FIR flux is certainly robust, and the slope of this relation 
for non-AGN galaxies less 
than 1, in agreement with the photometric relation between total infrared and
ISOCAM LW2 emission studied by \citet{chary01}. Applying such a 
luminosity-dependent ratio of PAH to FIR luminosity
to QSOs is consistent with our finding of star formation dominating the FIR 
emission in many of our QSOs. The alternate scenario, in which host galaxies
are all PAH luminous but FIR-quiescent and the far-infrared emission of 
QSOs is AGN dominated, is both complex and inconsistent with some 
of the evidence presented
in \S 3.2. Specifically, it cannot explain why both the far-infrared {\em and} 
the PAH emission rise together relative to the mid-IR continuum (i.e. AGN) 
luminosity (see also Figure~\ref{fig:plotpgpahnonpah}). Moreover, this 
scenario does not agree with the
multiwavelength evidence for active star formation mentioned in \S 4.3.

Considering these trends and the fact that the far-infrared luminosities
of the PG QSOs are typically in the $\gtrsim 10^{11}\Lsun$ regime, we 
conclude that for the average QSO in the sample the star formation 
contribution to the FIR emission
is at least 30\% (applying the starburst PAH to 60$\mu$m conversion), and that 
star formation may well be dominant. Comparing the locus of QSOs in 
Fig.~\ref{fig:pah_fir} with the trend for star-forming comparison objects 
suggests a tendency for the star formation contribution to be largest
in the most FIR-luminous QSOs.

 Similar considerations can 
be made for the comparison of \ne212m \ and far-infrared emission
in QSOs and ULIRGs (Fig.\ref{fig:ne2_fir}). Here, the 
sources of uncertainty are the NLR contribution to \ne212m \ for the QSOs
and the
possibility of different extinction in the QSO starbursts and the
ULIRGs which show considerable mid-IR extinction \citep{genzel98}.

\begin{figure}
\plotone{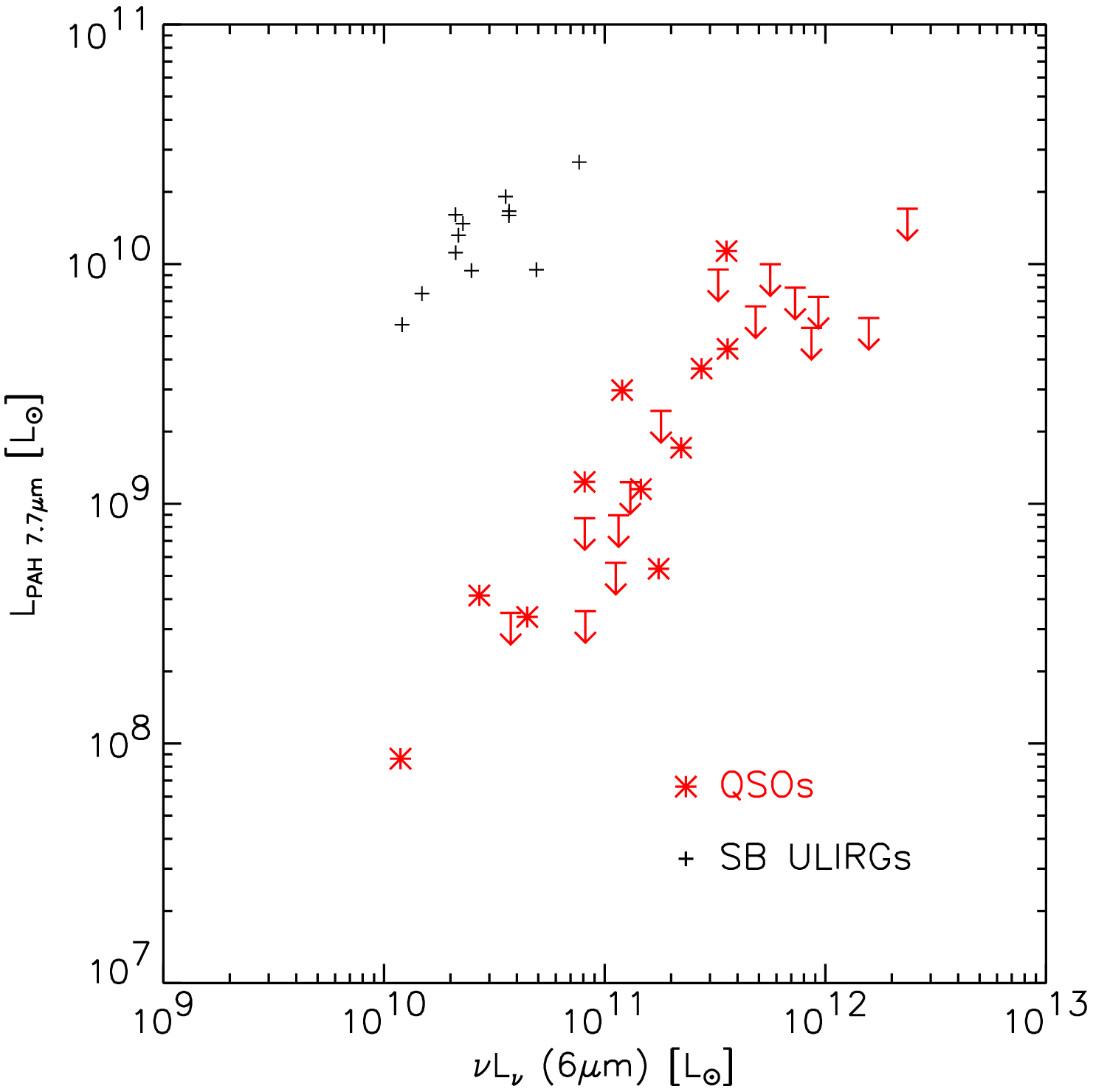}
\caption
{Luminosities L(\pahIR) vs. L(6$\mu$m) for the QSOs and for 
starburst dominated ULIRGs.}
\label{fig:pah_l6}
\end{figure}

The similarity between QSOs and starburst-dominated ULIRGs in trends and ratios
based on \pahIR, \ne212m , and far-infrared continuum no longer holds
when we plot a clearly AGN-related quantity on one axis. 
Fig.~\ref{fig:pah_l6} shows that
the ratio of PAH and 6$\mu$m continuum differs between QSOs and 
starburst-dominated ULIRGs by 
more than an order of magnitude, and that the two classes separate clearly
in the diagram. An additional strong non-starburst component is required in
the QSOs which clearly is the AGN-heated warm dust continuum. We note, 
however, a clear correlation between the luminosities of AGN 6$\mu$m 
continuum and starburst PAH among the QSOs. We will now argue on the basis 
of a more comprehensive 
comparison that this correlation is indirect, caused by a starburst-AGN
connection in QSOs.

In a flux-limited sample selected based on an emission component 
that is directly due to the AGN, as is the
case for the PG sample (selected by pointlike appearance, B magnitude and
blue U-B color), we expect correlations between the luminosities of 
the various AGN tracers to arise directly as a consequence of the sample 
selection. Correlations are not expected when comparing 
AGN- and starburst-related quantities unless there is a real 
physical correlation. 

\begin{figure}
\plotone{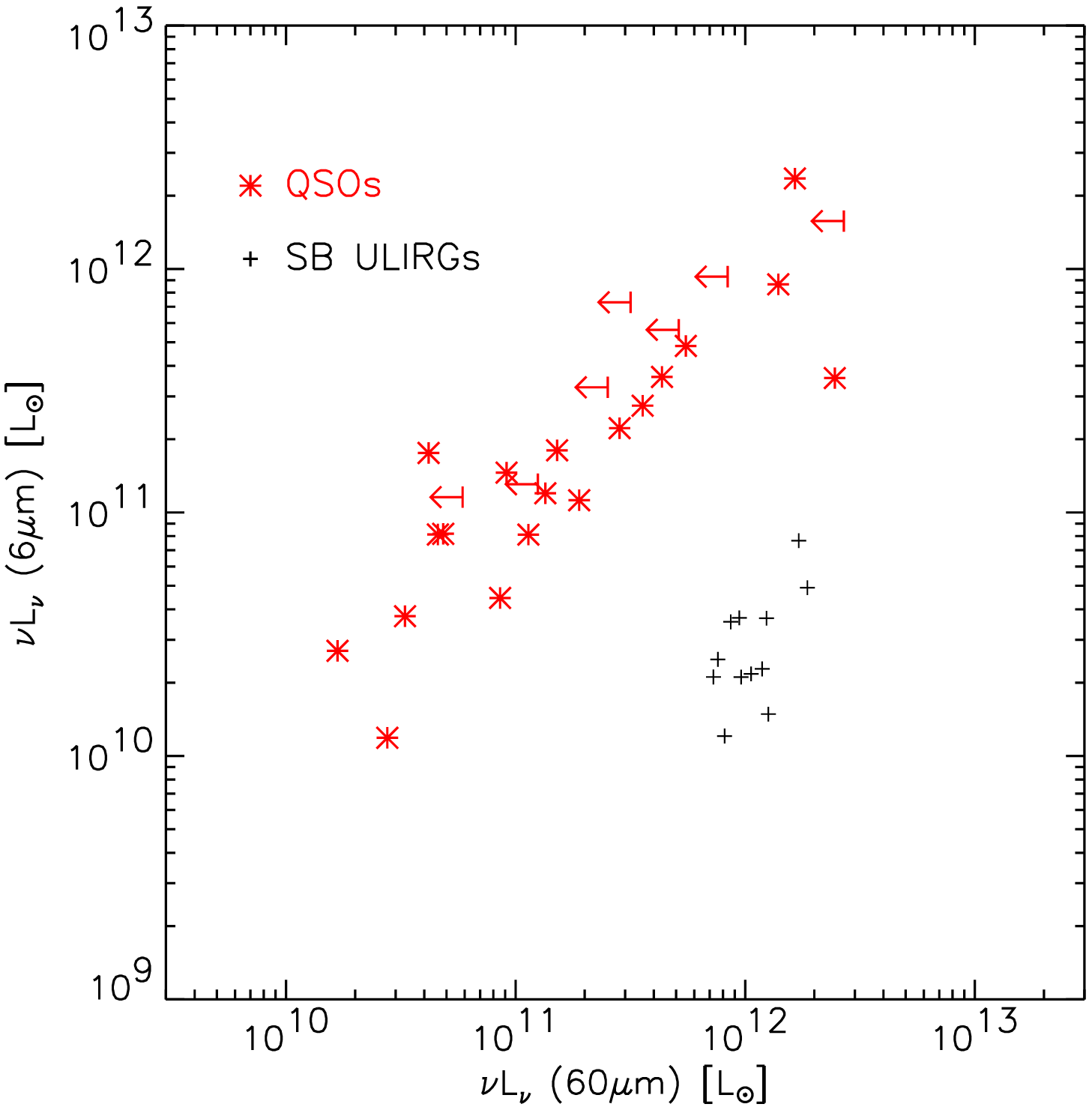}
\caption
{Luminosities L(6$\mu$m) vs. L(60$\mu$m) for the QSOs and for 
starburst dominated 
ULIRGs.}
\label{fig:l6_fir}
\end{figure}

\begin{figure}
\plotone{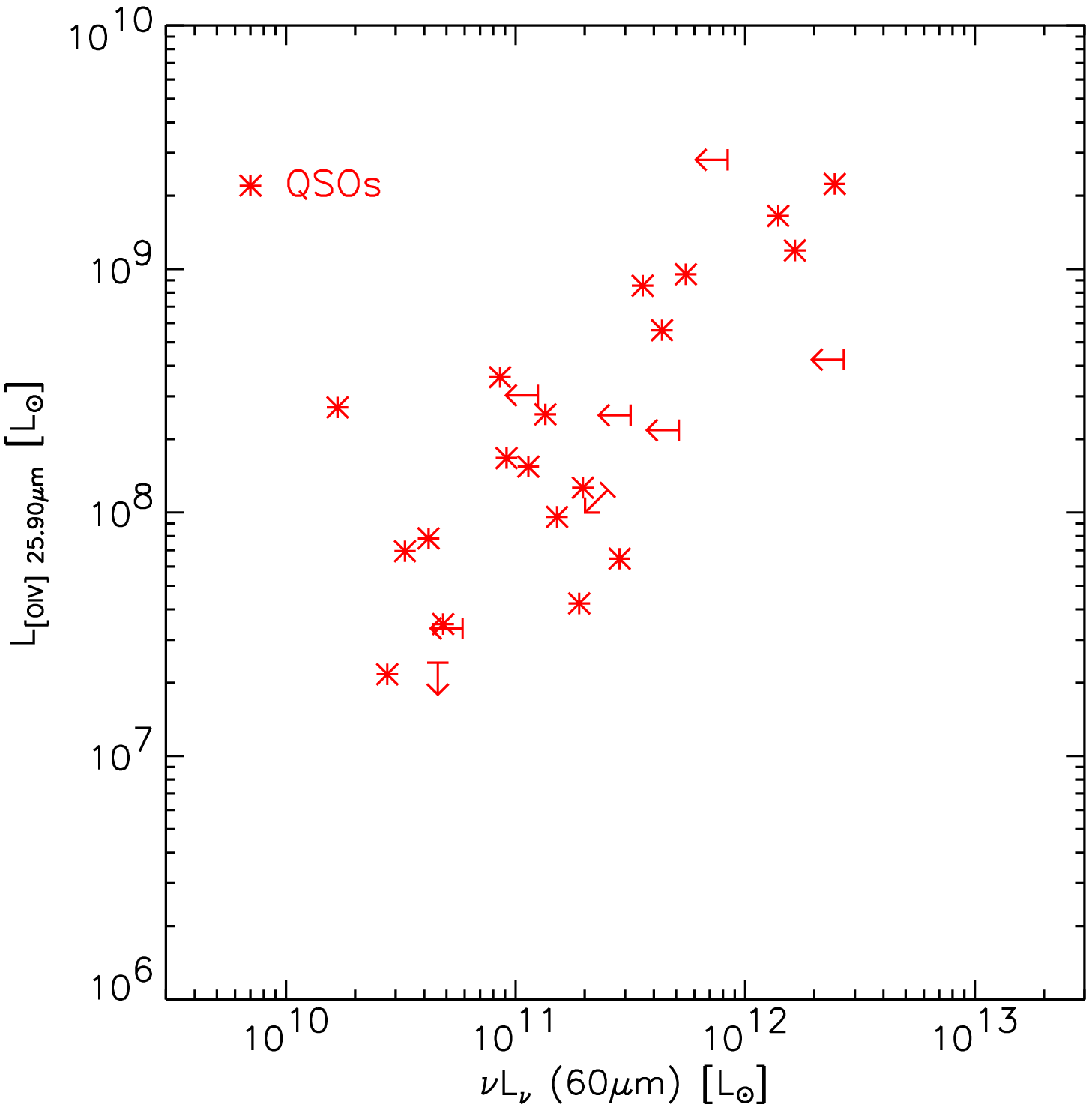}
\caption
{Luminosities L(\oivIR) vs. L(60$\mu$m) for the QSOs}
\label{fig:o4_fir}
\end{figure}

A summary of correlation coefficients among the continua, PAH and 
line fluxes measured in the QSOs is given in Table~\ref{tab:correlations}.
Some of the 
relations are also shown in Figs.~\ref{fig:pah_fir} to \ref{fig:o4_fir}. 
In Table~\ref{tab:correlations} we list Spearman rank correlation 
coefficients for the detected objects and their 
significance both for luminosities and for the observed fluxes.
Given the selection of the PG sample over a relatively narrow flux range,
the correlations in flux are generally less tight, but agree with the
findings based on luminosities. Our dataset includes upper limits, in 
particular for feature measurements and far-infrared fluxes, as well as 
trends between luminosity 
and distance caused by the flux limited selection. For these reasons, we also 
list in Table~\ref{tab:correlations} partial correlation coefficients 
(explicitly excluding the effect of distance) that have been 
computed using the 
formalism of \citet{akritas96} which provides partial correlation 
analysis for censored data. The results from this analysis of the
full censored dataset are in agreement with those from the detections only.

A first 
and expected finding is the significant correlations between various
luminosities tracing the AGN components, e.g. 6$\mu$m continuum and the 
high excitation emission lines. There is a noticeable spread in the ratio 
of mid-infrared continuum emission and emission in the mid-infrared high 
excitation lines \oivIR\ and \nevIR , however, equivalent to a spread in 
equivalent width of these 
lines. This is not uncommon in other AGN narrow emission lines 
like \oiii\  \citep[e.g.][]{boroson92,baskin05} and has implications
for the reliability of NLR lines as direct tracers of AGN bolometric 
luminosity. 
\citet{netzer06} discuss this large spread of \oiii\ equivalent
width which is likely a general property of AGN and also a function of
source luminosity (the `Baldwin-effect') and emission line reddening. 
 
Clear correlations are also seen between quantities tracing starburst 
activity (\pahIR\ being the cleanest) and 
others that in the QSOs must be almost fully dominated by the AGN 
(like 6$\mu$m continuum or high excitation lines).  
Such correlations are not caused by the PG sample selection by B-band flux
and U-B color,
and must indicate a true relation of more luminous QSOs on average
being associated with more luminous star formation. In the presence of
such a ``starburst-AGN connection'', evidence on the causal links connecting
the correlated observables has
to be obtained from the quality of the correlations, and in particular from 
comparing the absolute values of the correlated quantities with templates
as done above for far-infrared and PAH.

While the size of the present sample is modest, 
Table~\ref{tab:correlations} shows that the correlation between \pahIR\ and
far-infrared luminosities is one of the tighter among the combinations 
we investigated. Consistent with this trend is the stronger correlation of
\ne212m \ with far-infrared than is the case for far-infrared with
either \oivIR\ or \nevIR . \ne212m \ is a line
that is emitted both in the NLR of AGN \citep{sturm02}, with higher
excitation lines usually being stronger, and as the strongest mid-infrared
line in the spectra of most starbursts \citep{thornley00,verma03}. The 
good correlation of \ne212m \ and far-infrared thus indicates a strong 
starburst contribution to both  \ne212m \ and far-infrared emission, 
reinforcing the conclusion reached from the starburst-like ratios of these 
quantities. In a larger sample, these considerations could be expanded to a 
more rigorous test on the basis of the quality of the correlations. Even in 
the presence of a starburst-AGN connection, starburst tracers should usually 
correlate more tightly with other starburst tracers than with AGN tracers.
Our sample is not big enough for robust conclusions of this type. Looking at 
the probability of exceeding the partial correlation coefficients for the 
full censored dataset in the null hypothesis of uncorrelated data 
(column 11 of Table~\ref{tab:correlations}), it is nevertheless reassuring 
but certainly tentative
that the seven least significant of the 14 correlations discussed are 
60$\mu$m vs. 6$\mu$m, 60$\mu$m vs. \oivIR , 60$\mu$m vs. \nevIR , 
PAH vs. 6$\mu$m, PAH vs. \oivIR , PAH vs. \nevIR , 6$\mu$m vs. \nevIR . With 
the exception of the last, these less significant correlations are all of the
type starburst tracer vs. AGN tracer, in the superposition scenario outlined 
in \S 3.2, and provided that the far-infrared is counted as a star formation
tracer.

We conclude that, while there are also `indirect' correlations caused by
a global correlation of AGN and starburst luminosity in our PG sample, the
relation between PAH, [NeII], and far-infrared is real and reflects the
starburst component. The most important support for this interpretation 
is the starburst-like ratios of these three quantities. Starbursts contribute
at least $\sim$30\% and likely most of the far-infrared emission in the
average QSO in our sample. An upper limit to the starburst contribution
is imposed by the need for a realistic continuation to longer wavelengths
of the AGN mid-IR continuum, which cannot fall off more steeply than the 
Rayleigh-Jeans like
emission of optically thin dust of an appropriate temperature. The true slope 
is likely somewhat shallower due to variation in temperature and due
to non-negligible optical depth in part of the mid-IR emitting region. 
We feel that the origin and
interplay of silicate emission and continuum in the SED of the AGN
is not yet well enough measured or modelled for an accurate AGN continuum 
extrapolation of this type.  This is also due to the need for
unambiguous decomposition of starburst, silicate, and AGN continuum. 
Nevertheless, we consider the minimum pure AGN far-infrared continuum
required by the data to be consistent with our global conclusion 
from the \pahIR\ and \ne212m \ emission. 

\subsection{Mid-infrared diagnostics and the starburst-AGN connection in
QSOs}

Extending earlier ground-based work \citep[e.g.,][]{roche91}, 
\citet{genzel98} and \citet{laurent00} presented the first 
empirical versions of the tools now used to separate AGN-powered 
from starburst-powered infrared galaxies. The basis of these tools is 
that the intensity of the PAH features in starburst-powered 
systems traces the starburst's (far-infrared) luminosity, while
these features are easily destroyed by the strong and hard AGN radiation.
Our present study widens the scope of the \citet{genzel98} work by
focusing on star formation signatures in high luminosity bona-fide AGNs. 
We have used our QUEST QSO sample to look for three starburst 
signatures, strong PAH features, strong [NeII] lines and strong far-infrared 
continuum. Although two of those ([NeII] and far-infrared) can also 
partly originate in AGN environments, we have argued in \S4.1 that
the three quantities scale with each other and are tracing significant
star formation in most objects of our PG QSOs sample. The measured 
far-infrared luminosity
\nulnu60 \ ranges between $1.7\times 10^{10}$ and 2.5$\times 10^{12}\Lsun$, 
covering a wide range of starburst luminosity up to the ULIRG regime, and
limits for the remaining QSOs are consistent with starburst emission in 
the same range of luminosities.

An important result is the correlation of PAH (starburst) luminosity
and AGN luminosity in our sample. This extends to higher luminosity a similar
result obtained by S02 on the basis of \ISO\ spectroscopy of 
mostly lower luminosity Seyferts, and of a wide range of optical and
near-infrared studies
suggesting elevated starburst activity in Seyferts
\citep[e.g.][]{heckman97,oliva99,gonzalez01,imanishi03,kauffmann03}. 
Such a connection between small scale AGN feeding and larger scale 
starburst activity is plausible \citep[e.g.,][]{norman88}
and may play a role in establishing the 
black hole mass to bulge velocity dispersion relation in galaxies. Its 
details are far from trivial, however,  and warrant observations
with higher spatial resolution to elucidate the spatial
structure of star formation in these QSOs 
\citep[see for example][]{cresci04}.

Another effect of this connection relates to the interpretation of 
correlations between QSO properties measured at different wavelengths.
Some observed correlations may be indirect, driven by the 
starburst-AGN connection, as argued for example by \citet{roro95} for optical
and far-infrared continua.
\citet{haas03}, however, have used among other arguments the observed
 correlation between rest frame
mid-infrared and far-infrared continuum in their QSO sample
to argue for an AGN origin of the latter. The sensitivity of the \ISO\ 
spectroscopic data they had available did not allow for a conclusive test 
on the
basis of PAH emission. While we confirm the mid- to far-infrared correlation
for QSOs for our PG sample (e.g., Fig.~\ref{fig:l6_fir}), 
we use our higher sensitivity \spitzer\ \pahIR\ and 
\ne212m \ data to argue for an indirect nature of this correlation for the
luminosity range covered by our sample, induced by a starburst-AGN connection.
A similar test remains to be done for
the highest luminosity members of the \citet{haas03} sample, which are not
sufficiently represented in our local PG sample or in the IRS spectra
of radio galaxies and radio-loud QSOs of \citet{haas05}. Such high quality 
mid-infrared spectra of highest luminosity QSOs should also be able to test 
whether a trend for decreasing FIR to MIR ratio at highest optical luminosity 
\citep[as suggested by Fig.~4 of][]{haas03} reflects an increase in 
relative AGN intensity compared to the host and its star formation.
  
A central question in the study of QSOs and ULIRGs is their possible 
evolutionary relation \citep[e.g.][]{sanders88}. Our finding of luminous 
starburst activity in many QSOs is clearly consistent with such an evolutionary
link between ULIRGs (which show ultraluminous star formation and frequent 
coexisting AGN) and QSOs (which show ultraluminous AGN and frequent coexisting 
starbursts). From such basically energy-related considerations, an 
evolutionary path with a clear time arrow is, however, difficult to 
demonstrate and distinguish from more random processes. Including 
structural and dynamical information, e.g. from other elements of the 
QUEST program, will better probe this link. 

\subsection{Direct AGN heating of cold dust and PAHs?}

Several models have proposed a direct AGN heating of the far-infrared emission
of QSOs \citep[e.g.,][]{sanders89}. A basic feature of such models,
needed in order to fit QSO SEDs with moderately strong far-infrared emission,
is a significant covering factor by obscuring dust at relatively large
(few kpc) distances from the central AGN that is not shadowed by 
matter closer in. In the model of \citet{sanders89}, for example, this is 
accomplished by invoking a dusty galactic disk that is warped into
the unshielded AGN radiation on such scales.
In such a scenario, the star formation activity and associated PAH emission 
would be low. The PAH to far-infrared correlation that we find
would have to be due to PAH excitation by the AGN itself at this relatively 
large distance. The required emission is significant - as we argued above, 
the ratio of PAH and far-infrared is similar to that in 
starburst-dominated systems, 
and far-infrared is a significant fraction of the bolometric luminosity for 
our sample ($\sim$10\% mean for the systems with far-infrared detections). 

A main argument against this scenario is the likely destruction
mechanism of PAH by AGN radiation: If this process works through destruction
by individual EUV and X-ray photons, then the PAH carriers cannot survive 
at even kpc distances unless shielded by a large column able to stop the 
deeply penetrating
hard photons \citep{voit92}. Such a large absorbing column 
(N$\rm _H\gtrsim 10^{22}cm^{-2}$) would, however, prevent the heating of a 
significant far-infrared emitting dust component at large 
distance, by absorbing the UV `big blue bump' bulk part of the AGN SED. 
It would also absorb the near-UV ($<$13.6eV) radiation needed to actually 
excite infrared feature emission from the shielded PAH molecules. As argued by 
\citet{voit92} and \citet{maloney99}, PAH molecules can survive even 
relatively close to a powerful AGN if placed behind large obscuration.
Such high obscuring columns could be plausibly identified with 
the anisotropic obscuring structure 
postulated by unified AGN models. Exciting these surviving PAH molecules 
will however require a separate near-UV source,
for example by reintroducing a circumnuclear starburst. An indirect 
transport of AGN near-UV 
radiation to this shielded material, for example by scattering of UV emerging
into the AGN ionisation cones, appears 
unlikely to produce the required large PAH luminosities for $\sim$1\% 
scattering efficiency, as often assumed on the basis of polarimetric AGN
studies \citep[e.g.][]{pier94}.

\citet{freudling03} and \citet{siebenmorgen04b} present observations and 
models of radio-loud AGN in which they ascribe the full infrared SED 
including sometimes present PAH emission to the AGN, without invoking 
additional star formation. Their models 
\citep[see discussion in][]{siebenmorgen04a} invoke a different treatment of
PAH destruction in the AGN radiation field. While this issue certainly deserves
future study, their models may underestimate PAH destruction by the AGN. They 
predict strong PAH emission from optically thin dust illuminated by an AGN at 
a typical NLR distance, as well as significant PAH emission from an 
optically thick model for the central region of a nearby AGN which
has most of its mid-infrared emission arising in a compact ($\sim$10 pc) 
region \citep[Fig. 20, 21, 22 of][]{siebenmorgen04a}.
In contrast, spatially resolved observations put very strong limits on the
PAH emission from such regions of some nearby AGN 
\citep{lefloch01,siebenmorgen04a,weedman05,mason06}.

Our interpretation is also consistent with other, partly circumstantial,
evidence for star formation in QSOs. \citet{canalizo01} find optical 
spectroscopic evidence for relatively recent $\lesssim 300$Myr star formation
in a sample that ranges from ULIRGs to some of the moderately FIR-bright
PG QSOs of our sample. \citet{veilleux06} find some PG QSOs brighter than the 
H-band fundamental plane, possibly indicating circumnuclear star formation.
Molecular gas detections of PG QSOs \citep[e.g.][]{evans01,scoville03},
while not directly probing star formation, suggest sufficient material
to power the far-infrared emission if star formation is efficient.

\subsection{Comparison to QSO star formation estimates based on \oii }

Our finding that star formation activity is able to power the far-infrared 
emission is in contrast to the result of \citet{ho05}. For a PG-based sample
partly overlapping with our sample, he found insufficient
star formation from an analysis using the \oii\ line as a star formation 
tracer. Large and uncertain corrections for extinction of this tracer are 
likely the main contributor to this discrepancy.  

\citet{ho05} used \oii\ fluxes for his QSOs from the literature and 
the calibration of \citet{kewley04} for 
{\em extinction-corrected} \oii\ as a star formation indicator that is also a 
function of metallicity: SFR([O{\sc II}],Z). 
\citet{kewley04} base their work on a sample of nearby field galaxies 
with detailed integrated optical spectroscopy, and star formation rates 
spanning 4 orders of magnitude centered on $\sim 1\Msun$yr$^{-1}$. 
They demonstrate that \oii\ is a star formation indicator as good as H$\alpha$
over this range, provided extinction can be corrected for individual objects, 
and individual metallicities are known.

To apply this calibration to QSOs whose optical spectra are strongly 
dominated by AGN 
emission, \citet{ho05} made three assumptions: (1) Screen attenuation 
of A$_V$=1 towards the star forming regions, (2) metallicity twice solar, 
(3) one third of the \oii\ emission comes from star formation. Another 
implicit assumption is (4) aperture corrections to \oii\ can be ignored 
for the hosts of this QSO sample with median z$\sim$0.09 observed with 
2-5\arcsec\ optical spectroscopic apertures. Under these assumptions, 
\cite{ho05} 
infers star formation rates of at most 20 $\Msun$yr$^{-1}$ but often much 
less, and typically
an order of magnitude below those inferred by ascribing the far-infrared 
emission to star formation.

While all four assumptions contribute to the uncertainties of this approach, 
we believe that 
assumption (1) about the reddening to the star forming regions is the key to 
the systematic difference to our \spitzer\ results. As 
noted, e.g., by \citet{kewley04}, extinction is a strong function of 
intrinsic \oii\ luminosity. At the short wavelength of \oii\ where 
A$_\lambda\sim 1.5\times$A$_V$, increased dust 
extinction can relatively easily offset much of an increase in 
intrinsic line luminosity. Since the far-infrared luminosities of our 
quasars reach beyond $10^{11}\Lsun$, it is instructive to compare to the 
systematic optical spectroscopic study of a complete sample of 
luminous infrared galaxies by \citet{poggianti00}, which is
addressing objects with L$_{FIR}\sim 10^{11.5}\Lsun$ for our cosmology. 
More than half of their objects have 
so-called e(a) optical spectra, characterized by absorption in the higher 
Balmer lines, weak \oii , and significant obscuration of the emission lines. 
E(B-V) is $\sim$1.11 
even in the simplifying screen assumption, which still ignores regions 
in such 
highly dusty objects that are too obscured to contribute to the optical 
lines. With observed L(H$\alpha$)$\sim 10^{41}$erg s$^{-1}$ and 
\oii /H$\alpha\sim$0.23, they are in the regime of or below the portion (1/3)
of the QSO \oii\ luminosities that \citet{ho05} ascribes to star formation, 
already considering that \oii\ aperture corrections may be more significant for the \citet{poggianti00} sample which has median z=0.0324 and similar spectroscopic apertures.
Similar arguments apply, to a lesser degree, to the \citet{poggianti00} objects with spectral classifications other than e(a). There is a large population of dusty luminous starbursts that could fit the weak \oii\ emission of the 
\citet{ho05} QSOs. 

In summary, while the QSO optical data are consistent with the 
low star-formation rate interpretation of \citet{ho05}, they are also 
consistent with much higher star formation rates. This is because the 
optical analysis of \citet{ho05} is based on a single line that is
strongly extinction sensitive, and no extinction constraints are available 
for this component from the optical data. Less extinction sensitive data 
like the 
\spitzer\ infrared spectra are needed to break this degeneracy, and suggest
more substantial star formation rates. Some of the QSO host optical spectra 
may share properties of infrared galaxies with e(a) optical spectra: little 
obscuration towards the stellar continuum spectra dominated by an older 
post-burst component, but still significant obscuration of the active star 
forming regions \citep{poggianti00}.

\subsection{Implications for high redshift QSOs}

Deep submm and mm photometry has led to the detection of rest frame
submm and far-infrared dust emission from radio-quiet QSOs
at redshifts up to 6.42 \citep[e.g.][]{omont01,isaak02,bertoldi03}. It has
been variously argued through indirect arguments like CO content that this 
emission is star formation powered \citep[e.g.][]{walter03}, implying that 
these QSOs coexist with
extremely powerful $\gtrsim 10^{13}\Lsun$ starbursts. Our results support a
starburst origin of QSO far-infrared emission, but do not extend to this
luminosity and redshift regime. If the ratio of
PAH to far-infrared emission for these QSOs is similar to the one in 
the local QSOs and in ULIRGs, detection of PAH emission
on top of a strong continuum may be within reach of \spitzer\ spectroscopy.
PAH emission from similar luminosity SMGs is 
detectable at ULIRG-like ratios to the far-infrared emission \citep{lutz05}. 
Luminous PAH emission has also been reported in mid-infrared selected samples
of z$\sim$2 infrared galaxes \citep{yan05}. 
Probing for PAH emission may currently be the only
way to verify the assumption of simultaneous strong star formation and
QSO activity in high redshift QSOs.

\section{Conclusions}

Sensitive \spitzer\ mid-infrared spectroscopy reveals the widespread
presence of aromatic `PAH' emission features in z$\lesssim 0.3$ QSOs from the
Palomar-Green sample, indicating
the presence of powerful  (\nulnu60 $\sim 1.7\times 10^{10}$ to 
2.5$\times 10^{12}\Lsun$) star formation activity in these systems. 
Starburst and AGN activity are connected in QSOs up to these high luminosities.
By comparing the ratios of \pahIR, \ne212m , and far-infrared emission in QSOs
with starbursts we conclude that for the average QSO in our sample at 
least 30\% and likely most of the QSO far-infrared emission is due to star 
formation. The data suggest a trend with the star formation 
contribution being the largest in the most FIR-luminous QSOs.

\begin{acknowledgements}
We thank Aprajita Verma, Natascha F\"orster Schreiber and Helene Roussel
for help with comparison starburst data. We are grateful for comments by
the referee.

This work is based on observations carried out with the Spitzer Space
Telescope, which is operated by the Jet Propulsion Laboratory, California
Institute of Technology, under NASA contract 1407. Support for this work
was provided by NASA through contract 1263752 (S.V.) issued by
JPL/Caltech. The authors wish to recognize the very significant
cultural role that Theresienwiese has always had with the indigenous
Bavarian community.

HN acknowledges a Humboldt Foundation prize and thanks the host institution, 
MPE Garching, where this work has been done. AJB acknowledges support from 
the National Radio Astronomy Observatory, which is operated by Associated 
Universities, Inc., under cooperative agreement with the National Science
Foundation.  
\end{acknowledgements}

\newpage


%
\clearpage
\begin{deluxetable}{llrrrrrrrr}
\tabletypesize{\footnotesize}
\tablewidth{0pt}
\tablecaption{QSO sample\label{tab:targets}}
\tablehead{
\colhead{Object}&
\colhead{z}&
\colhead{S$_{6}$}&
\colhead{S$_{15}$}&
\colhead{S$_{30}$}&
\colhead{S$_{60}$}&
\colhead{S$_{100}$}&
\colhead{Ref}&
\colhead{D$_L$}&
\colhead{log($\rm\nu L_\nu$(60$\mu$m))}\\
\colhead{}&
\colhead{}&
\colhead{mJy}&
\colhead{mJy}&
\colhead{mJy}&
\colhead{mJy}&
\colhead{mJy}&
\colhead{}&
\colhead{Mpc}&
\colhead{\Lsun}\\
\colhead{(1)}&
\colhead{(2)}&
\colhead{(3)}&
\colhead{(4)}&
\colhead{(5)}&
\colhead{(6)}&
\colhead{(7)}&
\colhead{(8)}&
\colhead{(9)}&
\colhead{(10)}
}
\startdata
PG0026+129          &0.1420& 16.2& 35.8&  74&$<$162&$<$129&H00& 672&$<$11.10\\
PG0050+124 (IZw1)   &0.0611&178.3&515.1&1183&  2243&  2634&FSC& 274&   11.45\\ 
PG0157+001 (Mrk1014)&0.1630& 32.1&217.3&1352&  2224&  2164&FSC& 781&   12.39\\
PG0838+770          &0.1310& 12.1& 46.3& 105&   167&   180&N86,H03& 615&11.06\\
PG0953+414          &0.2341& 21.3& 33.0&  34&$<$129&$<$315&N86&1170&$<$11.71\\
PG1001+054          &0.1605& 16.8& 34.5&  69&   140&   146&H03& 768&   11.18\\
PG1004+130          &0.2400& 17.2& 74.8& 164&   191&$<$284&S89&1203&   11.74\\
PG1116+215          &0.1765& 54.5& 78.4& 113&$<$219&$<$285&H03& 853&$<$11.50\\
PG1126-041 (Mrk1298)&0.0600& 37.2&101.9& 311&   669&  1172&N86& 269&   10.93\\
PG1229+204 (Mrk771) &0.0630& 28.2& 88.3& 183&   241&   317&H03& 283&   10.52\\
PG1244+026          &0.0482& 15.8& 66.7& 194&   368&   362&H03& 214&   10.44\\
PG1302-102          &0.2784& 21.3& 80.4& 201&   343&   343&S89,H00&1425&12.14\\
PG1307+085          &0.1550&     & 50.2& 101&   212&   155&S89,H03& 739&11.29\\
PG1309+355          &0.1840& 22.1& 70.9& 106&$<$162&$<$192&H03& 893&$<$11.40\\
PG1411+442          &0.0896& 61.4& 96.9& 139&   147&   140&H00& 410&   10.62\\
PG1426+015          &0.0865& 55.1&135.3& 251&   350&   312&H03& 395&   10.96\\
PG1435-067          &0.1260& 18.3& 33.3&  77&   304&$<$333&H03& 590&   11.28\\
PG1440+356 (Mrk478) &0.0791& 55.1&135.3& 251&   597&   780&H03& 359&   11.13\\
PG1448+273          &0.0650& 19.0& 70.1& 117&   117&$<$252&S89& 292&   10.22\\
PG1613+658 (Mrk876) &0.1290& 55.7&120.2& 298&   591&  1002&H00& 605&   11.64\\
PG1617+175          &0.1124& 24.4& 41.0&  56& $<$98&$<$252&S89& 522&$<$10.77\\
PG1626+554          &0.1330& 11.7& 14.0&   3&$<$156&    70&H03& 626&   10.66\\
PG1700+518          &0.2920& 51.4&127.8&    &   348&   374&H03&1505&   12.22\\
PG2214+139 (Mrk304) &0.0658& 56.0& 76.8&  80&   337&$<$282&N86& 296&   10.68\\
B2 2201+31A         &0.2950& 32.2& 58.4&    &$<$295&$<$870&SP &1553&$<$12.43\\
PG2251+113          &0.3255& 15.4& 35.6&    & $<$67&$<$214&N86&1706&$<$11.92\\
PG2349-014          &0.1740& 21.2& 59.8& 167&   271&   290&S89& 840&   11.55\\
\enddata
\tablecomments{
Col. (1) --- Source name.\\
Col. (2) --- Redshift.\\
Cols. (3-5) --- Observed narrow band continuum flux densities around 6, 15, 
and 30 $\mu$m rest wavelength, extracted from the \IRS\ spectra. Missing 
entries are outside the rest wavelength range available for a given object.\\
Cols. (6-8) --- \ISO\ or \IRAS\ fluxes at observed 60 and 100 $\mu$m, and 
related references: FSC -- IRAS Faint Source Catalog, 
H00 -- \citet{haas00}, H03 -- \citet{haas03},
N86 -- \citet{neugebauer86}, S89 -- \citet{sanders89}, 
SP -- from IRAS Addscan/Scanpi\\
Col. (9) --- Luminosity distance in Mpc for a 
H$_0$=70km\,s$^{-1}$\,Mpc$^{-1}$, 
$\Omega_m=0.3$ and $\Omega_{\Lambda}=0.7$ cosmology\\
Col. (10) --- Far infrared luminosity. See text for treatment of limits.\\ 
}
\end{deluxetable}

%
\clearpage
\begin{deluxetable}{lrrrr}
\tabletypesize{\small}
\tablewidth{0pt}
\tablecaption{Measured emission features\label{tab:emission_lines}}
\tablehead{
\colhead{Object}&
\colhead{\ne212m }&
\colhead{\nevIR }&
\colhead{\oivIR }&
\colhead{\pahIR }\\
\colhead{}&
\colhead{W\,cm$^{-2}$}&
\colhead{W\,cm$^{-2}$}&
\colhead{W\,cm$^{-2}$}&
\colhead{W\,cm$^{-2}$}
}
\startdata
PG0026+129    &   2.29E-22&   4.73E-22&   2.14E-21&$<$8.70E-21\\
PG0050+124    &   1.94E-21&   5.50E-21&   2.75E-21&   7.29E-20\\
PG0157+001    &   5.52E-21&   5.18E-21&   1.17E-20&   5.93E-20\\ 
PG0838+770    &   4.11E-22&   3.24E-22&   1.30E-21&   1.04E-20\\
PG0953+414    &$<$1.70E-22&$<$1.90E-22&   5.08E-22&$<$2.33E-20\\
PG1001+054    &   4.00E-22&$<$1.20E-22&   5.19E-22&$<$1.32E-20\\
PG1004+130    &$<$2.43E-22&$<$2.76E-22&   2.10E-21&$<$1.47E-20\\
PG1116+215    &$<$3.20E-22&$<$2.90E-22&   1.10E-21&$<$3.50E-20\\
PG1126-041    &   1.39E-21&   4.34E-21&   1.59E-20&   1.49E-20\\
PG1229+204    &   6.13E-22&   9.06E-22&   2.77E-21&$<$1.40E-20\\
PG1244+026    &   9.42E-22&   5.31E-22&   1.51E-21&   6.02E-21\\
PG1302-102    &   3.56E-22&   4.87E-22&   2.60E-21&$<$8.52E-21\\
PG1307+085    &   3.98E-22&   5.63E-22&   7.38E-22&           \\
PG1309+355    &   5.07E-22&   2.69E-22&$<$4.95E-22&$<$3.80E-20\\
PG1411+442    &   3.61E-22&   9.56E-22&   1.49E-21&   1.02E-20\\
PG1426+015    &   1.29E-21&   1.25E-21&   3.43E-21&   2.36E-20\\
PG1435-067    &$<$1.05E-22&   5.22E-22&   3.88E-22&$<$5.20E-21\\
PG1440+356    &   4.11E-21&   1.33E-21&   6.26E-21&   7.38E-20\\
PG1448+273    &   5.07E-22&   2.67E-21&   1.01E-20&   1.55E-20\\
PG1613+658    &   3.88E-21&   1.13E-21&   4.89E-21&   3.86E-20\\
PG1617+175    &   2.89E-22&$<$1.70E-22&   3.92E-22&$<$1.05E-20\\
PG1626+554    &   6.91E-23&$<$6.90E-23&$<$1.97E-22&$<$7.10E-21\\
PG1700+518    &   1.21E-21&$<$2.30E-22&   1.68E-21&$<$2.40E-20\\
PG2214+139    &   2.26E-22&   2.70E-22&   1.27E-21&$<$1.30E-20\\
B2 2201+31A   &   9.64E-23&   5.31E-22&   5.62E-22&$<$7.89E-21\\
PG2251+113    &   1.69E-22&   4.90E-22&   3.08E-21&$<$8.00E-21\\
PG2349-014    &   1.44E-21&   7.05E-22&   3.87E-21&   1.66E-20\\
\enddata
\end{deluxetable}
 
%
\begin{deluxetable}{llccccccccc}
\rotate
\tabletypesize{\small}
\tablewidth{0pt}
\tablecaption{Correlation of measured properties\label{tab:correlations}}
\tablehead{
\colhead{Property A}&
\colhead{Property B}&
\colhead{Number}&
\colhead{$R_S$}&
\colhead{Probability}&
\colhead{$R_S$}&
\colhead{Probability}&
\colhead{Dispersion}&
\colhead{Number}&
\colhead{PKT}&
\colhead{Probability}\\
&
&
\colhead{Detect.}&
\colhead{Lumin.}&
\colhead{Lumin.}&
\colhead{Flux}&
\colhead{Flux}&
&
\colhead{All}
\\
\colhead{(1)}&
\colhead{(2)}&
\colhead{(3)}&
\colhead{(4)}&
\colhead{(5)}&
\colhead{(6)}&
\colhead{(7)}&
\colhead{(8)}&
\colhead{(9)}&
\colhead{(10)}&
\colhead{(11)}
}
\startdata
60$\mu$m &PAH      & 11  &   0.945  &  1.1E-5 &0.582&6.0E-2&0.293&26&0.248&5.7E-3\\  
60$\mu$m &$6\mu$m  & 19  &   0.894  &  2.4E-7 &0.567&1.1E-2&0.303&26&0.321&1.6E-2\\
60$\mu$m &$15\mu$m & 20  &   0.938  &  9.8E-10&0.746&1.6E-4&0.212&27&0.441&2.8E-4\\
60$\mu$m &$30\mu$m & 19  &   0.926  &  1.3E-8 &0.825&1.4E-5&0.384&24&0.530&4.6E-5\\
60$\mu$m &[NeII]   & 18  &   0.899  &  4.0E-7 &0.744&4.0E-4&0.284&27&0.329&1.1E-3\\
60$\mu$m &[OIV]    & 19  &   0.670  &  1.7E-3 &0.493&3.2E-2&0.422&27&0.267&1.4E-2\\
60$\mu$m &[NeV]    & 16  &   0.788  &  2.9E-4 &0.456&7.6E-2&0.345&27&0.257&3.1E-2\\
PAH      &$6\mu$m  & 11  &   0.873  &  4.6E-4 &0.473&1.4E-1&0.283&26&0.174&6.2E-2\\
PAH      &$15\mu$m & 11  &   0.927  &  1.0E-5 &0.800&3.1E-3&0.219&26&0.277&5.3E-3\\
PAH      &$30\mu$m & 11  &   0.927  &  4.0E-5 &0.636&3.5E-2&0.268&23&0.346&1.6E-3\\
PAH      &[NeII]   & 11  &   0.936  &  2.2E-5 &0.845&1.1E-3&0.246&26&0.287&1.1E-3\\
PAH      &[OIV]    & 11  &   0.600  &  5.1E-2 &0.436&1.8E-1&0.397&26&0.186&2.6E-2\\
PAH      &[NeV]    & 11  &   0.682  &  2.1E-2 &0.636&3.5E-2&0.352&26&0.214&1.2E-2\\
$6\mu$m  &[OIV]    & 24  &   0.657  &  4.8E-4 &0.147&4.9E-1&0.482&26&0.264&8.4E-3\\
$6\mu$m  &[NeV]    & 19  &   0.770  &  1.2E-4 &0.484&3.6E-2&0.366&26&0.256&4.0E-2\\
\enddata
\tablecomments{
Col. (1) --- First variable. Wavelengths stand for continuum at that
rest wavelength.\\
Col. (2) --- Second variable.\\
Col. (3) --- Number of sources detected in both quantities.\\
Col. (4) --- Spearman's rank correlation coefficient for luminosities, for detected 
             sources.\\
Col. (5) --- Probability of exceeding the measured 
correlation coefficient for luminosities in the null hypothesis of 
uncorrelated data. Smaller values indicate more significant correlations.\\
Col. (6) --- Spearman's rank correlation coefficient for fluxes.\\
Col. (7) --- Probability of exceeding correlation coefficient for fluxes in
the null hypothesis.\\
Col. (8) --- Dispersion of Log$_{10}$(A/B) for the detected sources.\\
Col. (9) --- Total number of sources measured (detections and limits).\\ 
Col. (10) --- Partial Kendall $\tau$ coefficient describing correlation
between luminosities A and B, excluding the effect of distance. 
Value computed using the formalism of
\cite{akritas96} that extends the Kendall $\tau$-coefficient to partial 
correlation in the presence of censored data (i.e. data including upper 
limits).\\
Col. (11) --- Probability of exceeding partial correlation coefficient
in the null hypothesis.
}
\end{deluxetable}


\end{document}